\documentclass[aps,pra,twocolumn,showpacs,floatfix,superscriptaddress]{revtex4-1}

  \usepackage[utf8]{inputenc}
  \usepackage[T1]{fontenc}
  \usepackage{float} 
  \usepackage{color}  

\usepackage{graphicx}
\usepackage{amsmath}
\usepackage{amssymb}
\usepackage{bbm}
\usepackage{indentfirst}
\usepackage{dcolumn}
\usepackage{soul}

\begin{document}

\title{Complex magnetic phase diagram and skyrmion lifetime in an ultrathin film from atomistic simulations}

\author{Levente R\'{o}zsa}
\email{rozsa@phy.bme.hu}
\affiliation{Department of Theoretical Physics, Budapest University of Technology and Economics, Budafoki \'{u}t 8, H-1111 Budapest, Hungary}
\affiliation{Institute for Solid State Physics and Optics, Wigner Research Center for Physics, Hungarian Academy of Sciences,
P.O. Box 49, H-1525 Budapest, Hungary}
\author{Eszter Simon}
\author{Kriszti\'{a}n Palot\'{a}s}
\affiliation{Department of Theoretical Physics, Budapest University of Technology and Economics, Budafoki \'{u}t 8, H-1111 Budapest, Hungary}
\author{L\'{a}szl\'{o} Udvardi}
\author{L\'{a}szl\'{o} Szunyogh}
\affiliation{Department of Theoretical Physics, Budapest University of Technology and Economics, Budafoki \'{u}t 8, H-1111 Budapest, Hungary}
\affiliation{MTA-BME Condensed Matter Research Group, Budapest University of Technology and Economics, Budafoki \'{u}t 8, H-1111 Budapest, Hungary}
\date{\today}

\begin{abstract}

We determined the magnetic $B-T$ phase diagram of PdFe bilayer on Ir$(111)$ surface by performing Monte Carlo and spin dynamics simulations based on an effective classical spin model. The parameters of the spin model were determined by \textit{ab initio} methods. At low temperatures we found three types of ordered phases, while at higher temperatures, below the completely disordered paramagnetic phase, a large region of the phase diagram is associated with a fluctuation-disordered phase. Within the applied model, this state is characterized by the presence of skyrmions with finite lifetime. According to the simulations, this lifetime follows the Arrhenius law as a function of temperature.

\end{abstract}

\maketitle

\section{Introduction}

Stable localized field configurations in nonlinear field theory were identified by Skyrme in 1961\cite{Skyrme,Skyrme2}. It was shown later that such configurations may form thermodynamically stable states in magnetic systems\cite{Bogdanov,Bogdanov2}. 
These so-called magnetic skyrmion lattices (SkL) generally appear due to the Dzyaloshinsky--Moriya interaction\cite{Dzyaloshinsky,Moriya} (DMI) present in non-centrosymmetric systems. The identification of the A phase of the cubic B20 system MnSi as a physical realization of the skyrmion lattice\cite{Muhlbauer} increased interest in finding further materials with similar spin configurations. In the last few years, skyrmion states were also observed experimentally in bulk systems and thin films of FeGe\cite{Yu2,Wilhelm} and FeCoSi\cite{Munzer,Yu}; in Fe monolayer\cite{Heinze} and PdFe bilayer\cite{Romming} on Ir(111); in bulk Cu$_{2}$OSeO$_{3}$\cite{Adams} and GaV$_{4}$S$_{8}$\cite{Kezsmarki}; and in Pt|Co|Ir multilayer\cite{Moreau-Luchaire}. Since skyrmions are stable spin configurations which can be moved around by spin-polarized current, these systems have promising applications for spintronics devices\cite{Fert,Crum}.

Previous spin-polarized scanning tunneling microscopy (SP-STM) experiments performed on PdFe bilayer on Ir$(111)$ surface\cite{Romming,Romming2} concentrated on the low-temperature magnetic behavior of the system, up to $T=8\,\textrm{K}$. Concomitantly, the theoretical investigations\cite{Dupe,Simon,Crum} determined the parameters of a spin model Hamiltonian and discussed the ground state of the system as a function of the magnitude of the external magnetic field $B$. Three different phases were identified: a cycloidal spin spiral state (SS) with zero net magnetization for low fields, a ferromagnetic or field-polarized state (FP) for high $B$, and a hexagonal SkL state between them. This is in good agreement with previous theoretical calculations\cite{Bogdanov2,Han,Butenko} as well as Monte Carlo simulations (MCS) for similar anisotropic systems\cite{Yu,Yi,Polesya}. The experiments\cite{Romming,Romming2} also identified spin spirals, skyrmions and field-polarized domains, but the exact phase boundaries were not determined, probably due to the coexistence of metastable phases.


The $B-T$ phase diagram of MnSi has been extensively studied  experimentally\cite{Muhlbauer,Bauer,Bauer2,Bauer3,Janoschek,Zhang,Demishev,Wilson} and by MCS\cite{Buhrandt}. In this system, the SkL is only stable at finite temperature and magnetic field, in a small pocket of the phase diagram. At lower temperatures, this state turns into a conical SS state. The ordering temperature $T_{\textrm{c}}\approx 29\,\textrm{K}$ of the SS and SkL states only weakly depends on the magnetic field. Above $T_{\textrm{c}}$, there is a narrow ($1-2\,\textrm{K}$) transition region, which shows critical fluctuations based on neutron scattering\cite{Grigoriev,Pappas}, susceptibility and specific heat measurements\cite{Stishov,Bauer,Bauer2,Bauer3}. The theoretical description in Refs.~\cite{Janoschek,Bauer3,Buhrandt} identifies this behavior with a fluctuation-induced first-order phase transition proposed by Brazovski\u{i}\cite{Brazovskii}, where the fluctuations are isotropically strong on a surface of a sphere in momentum space. Earlier papers suggested\cite{Rossler,Pappas,Hamann} that the transition region may be attributed to the appearance of stable skyrmions in the system (even at zero external field), however, these skyrmions do not show long-range order as evidenced by the neutron scattering experiments.

In this paper, we determine the $B-T$ phase diagram of PdFe bilayer on Ir$(111)$ using Metropolis MCS and stochastic Landau--Lifshitz--Gilbert spin dynamics\cite{Landau,Gilbert,Brown,Kubo} simulations (SDS), based on a classical spin Hamiltonian
\begin{eqnarray}
H=-\frac{1}{2}\sum_{i \ne j}\boldsymbol{S}_{i}\mathcal{J}_{ij} \boldsymbol{S}_{j}-\sum_{i}m\boldsymbol{S}_{i}\boldsymbol{B},\label{eqn1}
\end{eqnarray}
where $\boldsymbol{S}_{i}$ is the spin unit vector at site $i$, $\mathcal{J}_{ij}$ denote tensorial coupling coefficients\cite{Udvardi}, $m$ is the magnetic moment of the Fe atoms and $\boldsymbol{B}$ is the external magnetic field.
The parameters $\mathcal{J}_{ij}$ and $m$ were determined from \textit{ab initio} calculations, reported in detail in Ref.~\cite{Simon} and summarized in Appendix~\ref{secS1}. Based on these simulations we demonstrate the presence of the strongly fluctuating intermediate region between the non-collinear ordered states and the paramagnetic state, displaying the characteristics of the fluctuation-disordered regime in MnSi\cite{Janoschek,Bauer3}. We calculate the finite lifetime of skyrmions in this region and stress the importance of different timescales available in experiments and simulations.

\section{Phase diagram from Monte Carlo simulations}

\begin{figure}
\includegraphics[width=0.9\columnwidth]{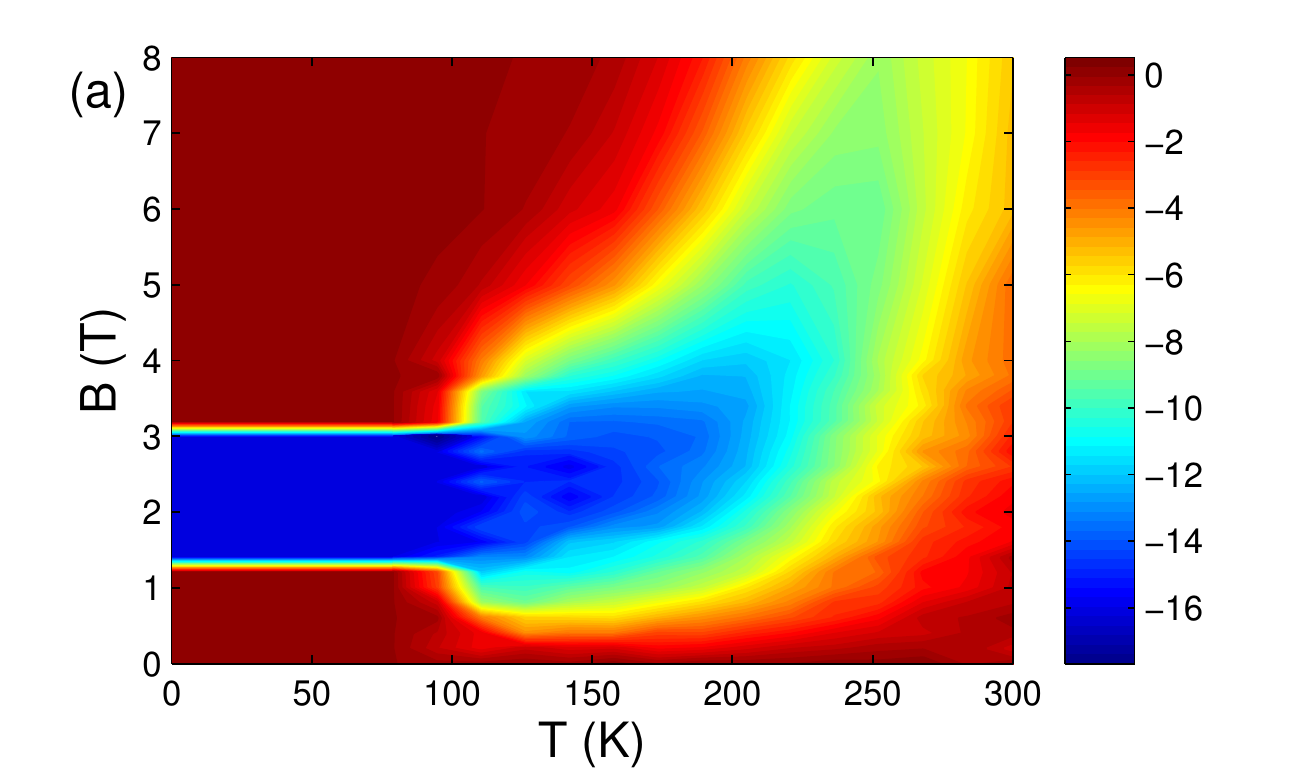}
\includegraphics[width=0.9\columnwidth]{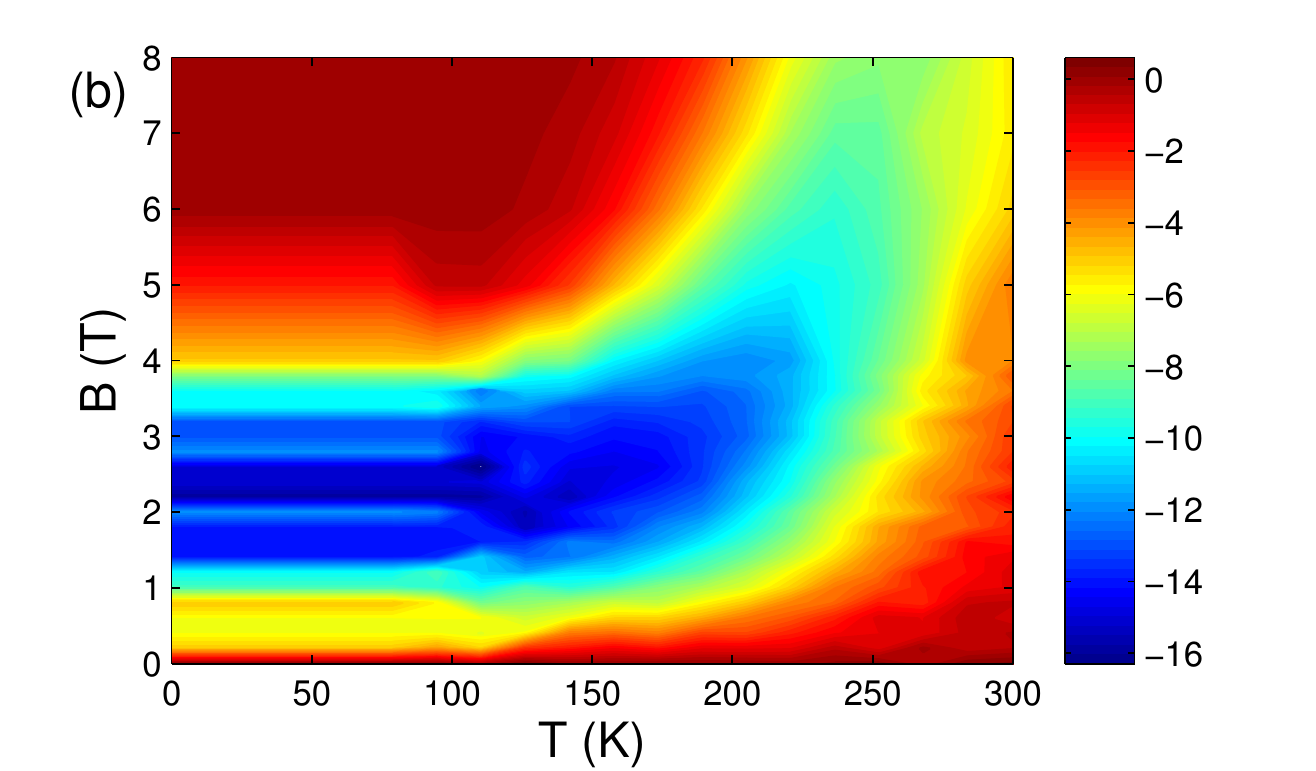}
\includegraphics[width=0.9\columnwidth]{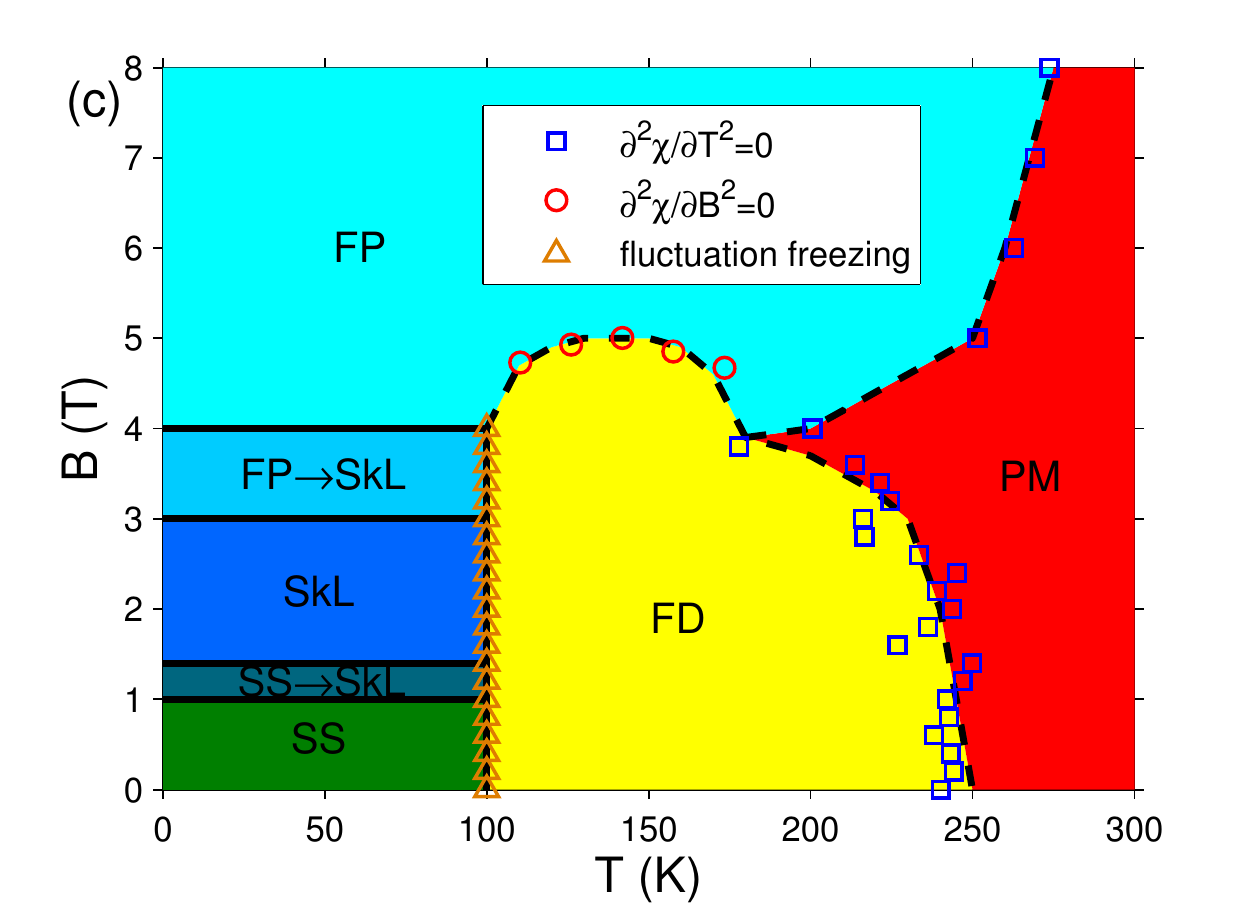}
\caption{(color online) (a)-(b)~Average topological number $Q$ in the system as a function of external field and temperature, calculated from MCS at a fixed magnetic field for (a) increasing and (b) decreasing temperature. Below $T_{\textrm{c}}\approx100\,\textrm{K}$, the topological charge did not change during the simulation time. (c)~Phase diagram constructed from the inflection points of the magnetic susceptibility, the line where the fluctuations of the system are frozen and the calculated ground states at zero temperature. Solid lines denote first-order transitions while staggered lines denote second-order transitions. The shallow regions between the ordered SS, SkL, and FP phases denote transitions which could not be resolved by the simulations.}
\label{fig2}
\end{figure}

We characterized the different phases in the $B-T$ plane by the average topological charge $Q$, the static structure factor $S\left(\boldsymbol{q}\right)$ and the static magnetic susceptibility $\chi$ -- see Appendices~\ref{secS2}-\ref{secS3} for the definitions. The simulations were performed on an $N=128\times128$ lattice with periodic boundary conditions; the effect of the boundary conditions on the SS and SkL phases is discussed in Appendix~\ref{secS4}. The average topological charge as a function of $B$ and $T$ is shown in Figs.~\ref{fig2}(a)-(b), for simulations performed at a fixed $B$ and by (a) increasing or (b) decreasing the temperature. The external field was oriented outwards from the Ir surface, which favors skyrmions with topological charge $Q=-1$ in the system as discussed in Appendix~\ref{secS3}. At low temperature, we identified three possible ordered states\cite{Bogdanov2,Han,Butenko} in this uniaxial system: cycloidal SS for $B \lessapprox1.4\,\textrm{T}$ , hexagonal SkL for $1.4\,\textrm{T}\lessapprox B \lessapprox3\,\textrm{T}$, and FP state for $3\,\textrm{T} \lessapprox B$, with the transition values of $B$ determined from energy minimization at zero temperature\cite{Simon}. The obtained field values are in good agreement with the ones reported in Ref.~\cite{Romming2}, although the exact phase boundaries are not reported there, probably due to the coexistence of the phases. The diameter $d_{0}$ of the circle around the skyrmion where the spins lie in the plane changes from approximately $3\,\textrm{nm}$ at $B=1.4\,\textrm{T}$ to $2\,\textrm{nm}$ at $B=3\,\textrm{T}$ in Ref.~\cite{Romming2}, while in our simulations it changed from $5.1\,\textrm{nm}$ to $3.5\,\textrm{nm}$ at the same field values. This indicates a relatively good agreement of the \textit{ab initio} calculations with the experiments, especially taking into account the strong dependence of the skyrmion size on the relaxation of the Fe layer with respect to the top Ir layer\cite{Simon}.

\begin{figure*}
\includegraphics[width=0.66\columnwidth]{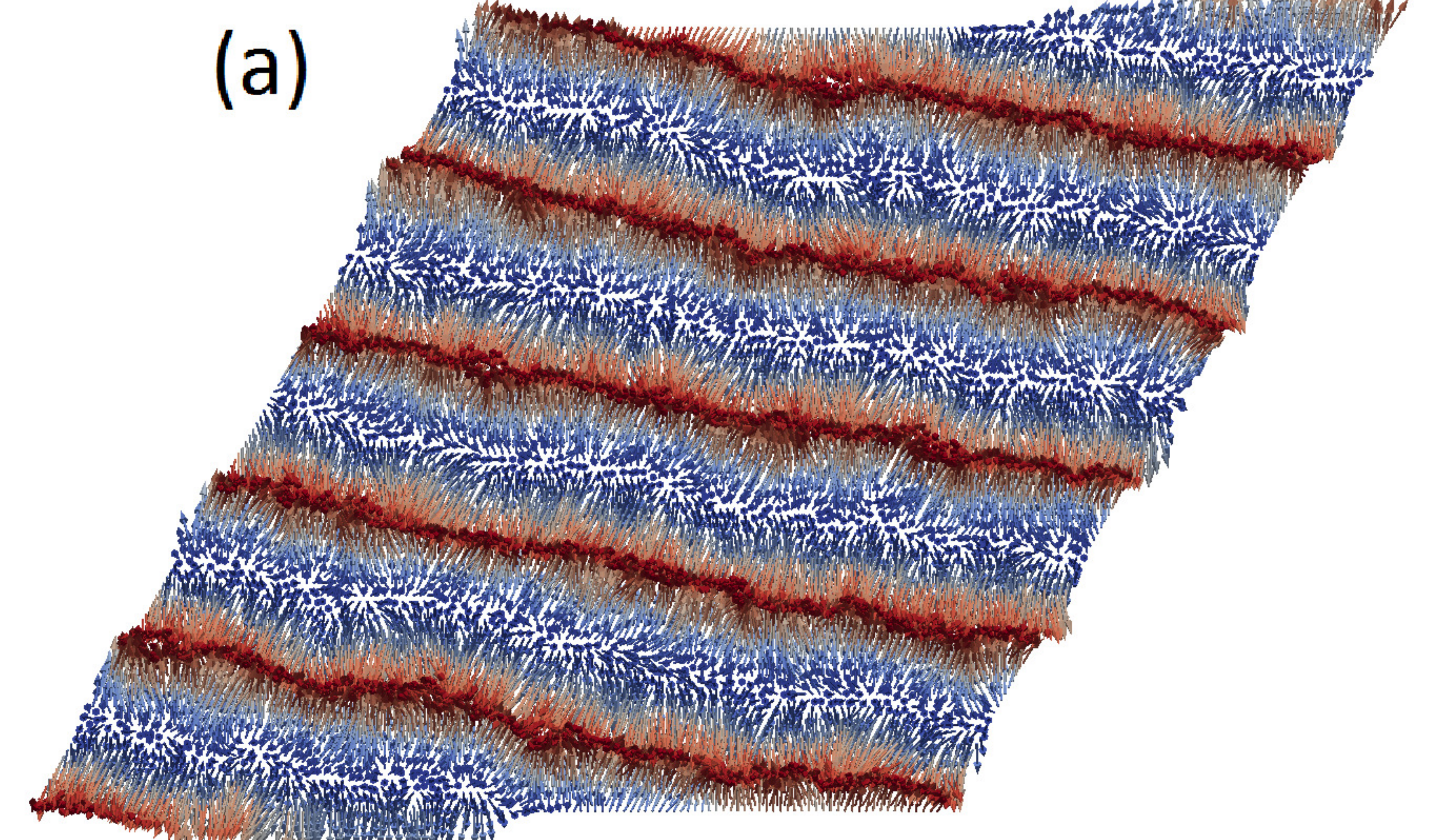}
\includegraphics[width=0.66\columnwidth]{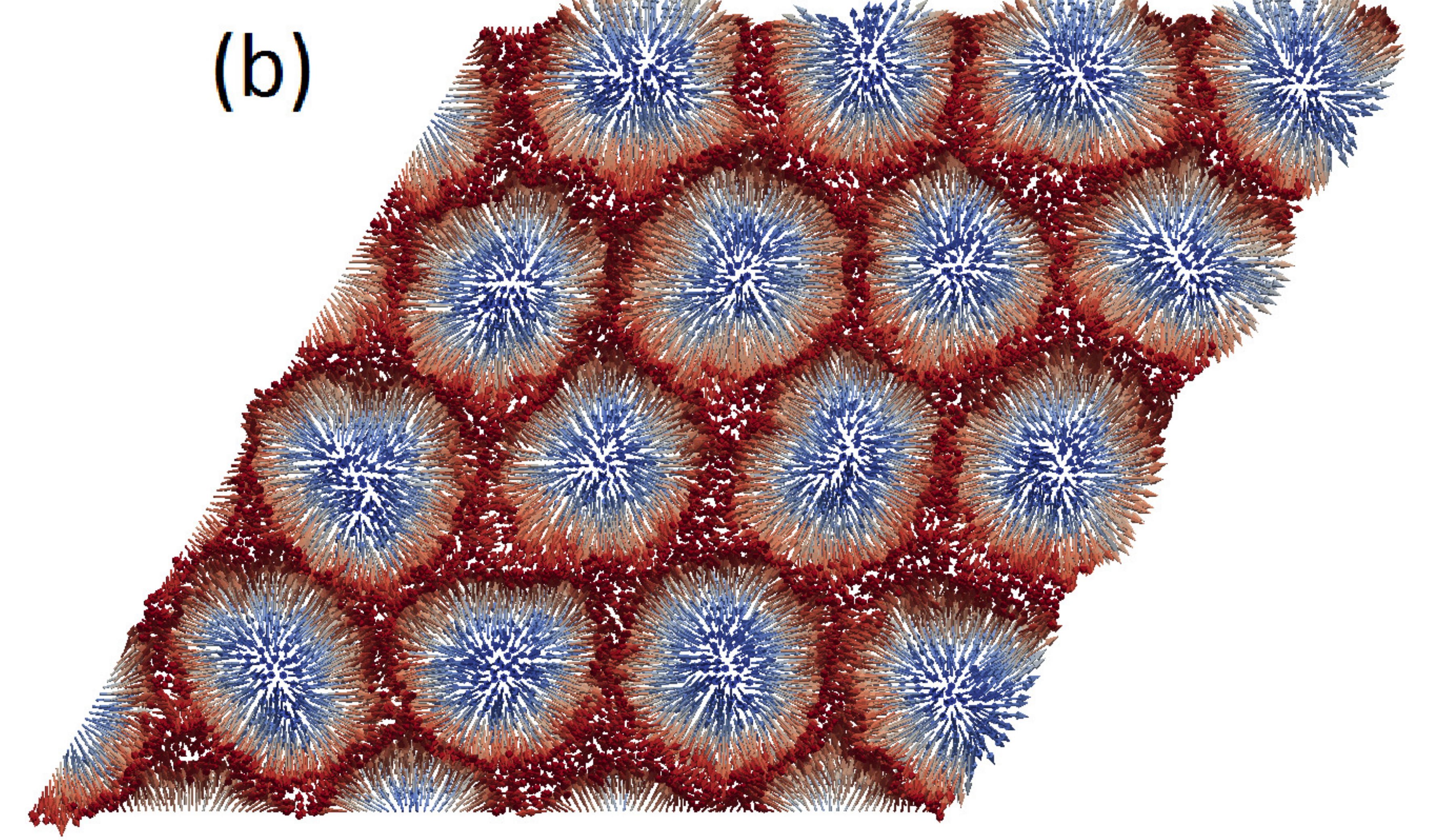}
\includegraphics[width=0.66\columnwidth]{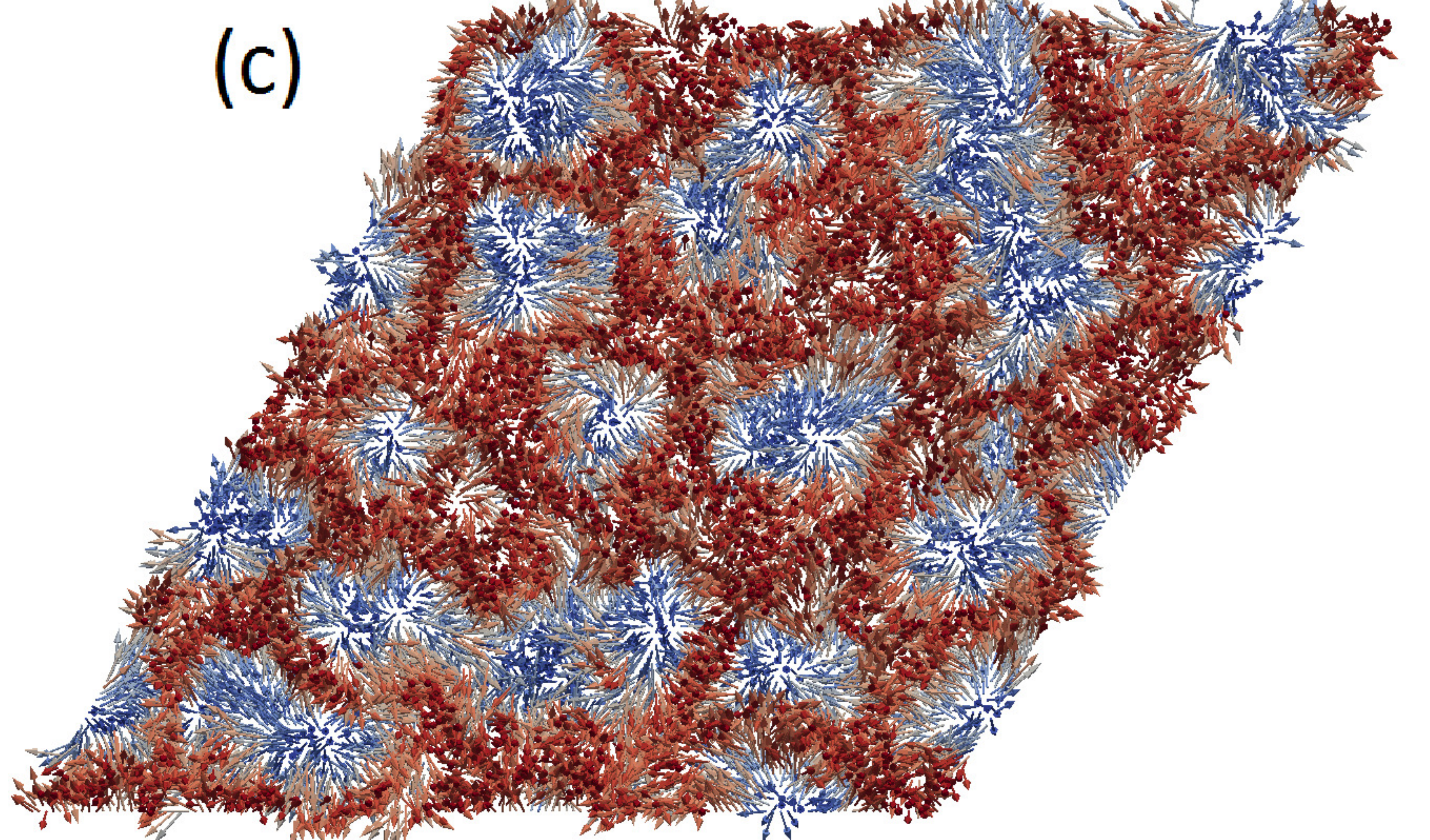}
\includegraphics[width=0.66\columnwidth]{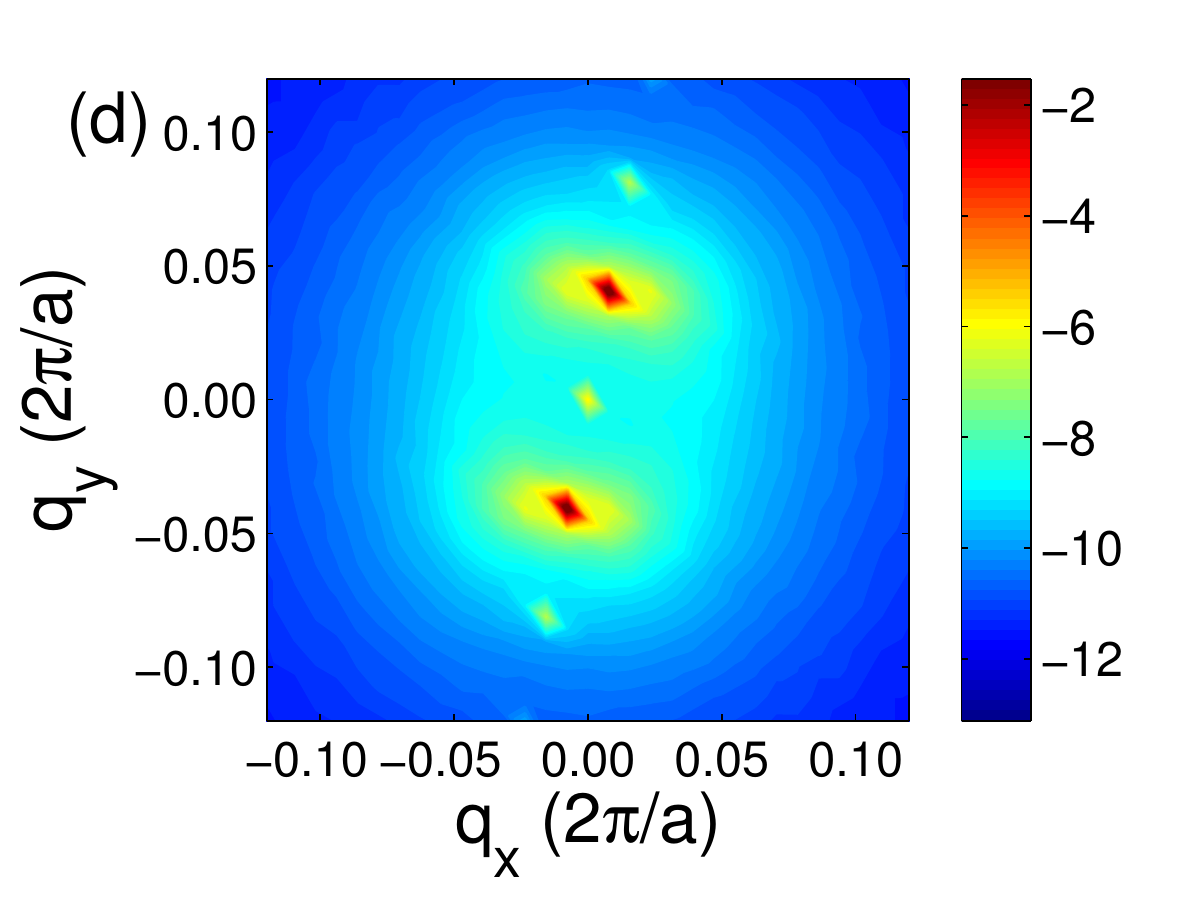}
\includegraphics[width=0.66\columnwidth]{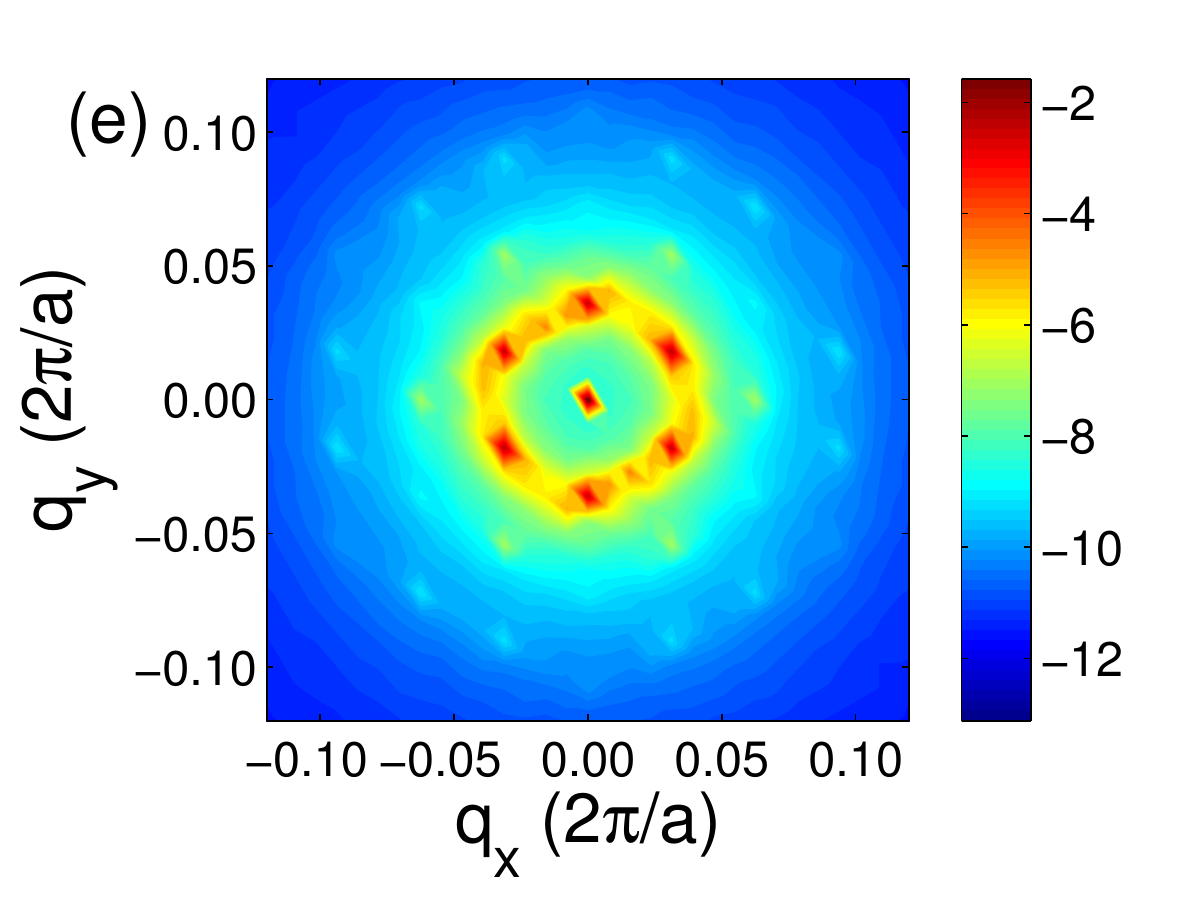}
\includegraphics[width=0.66\columnwidth]{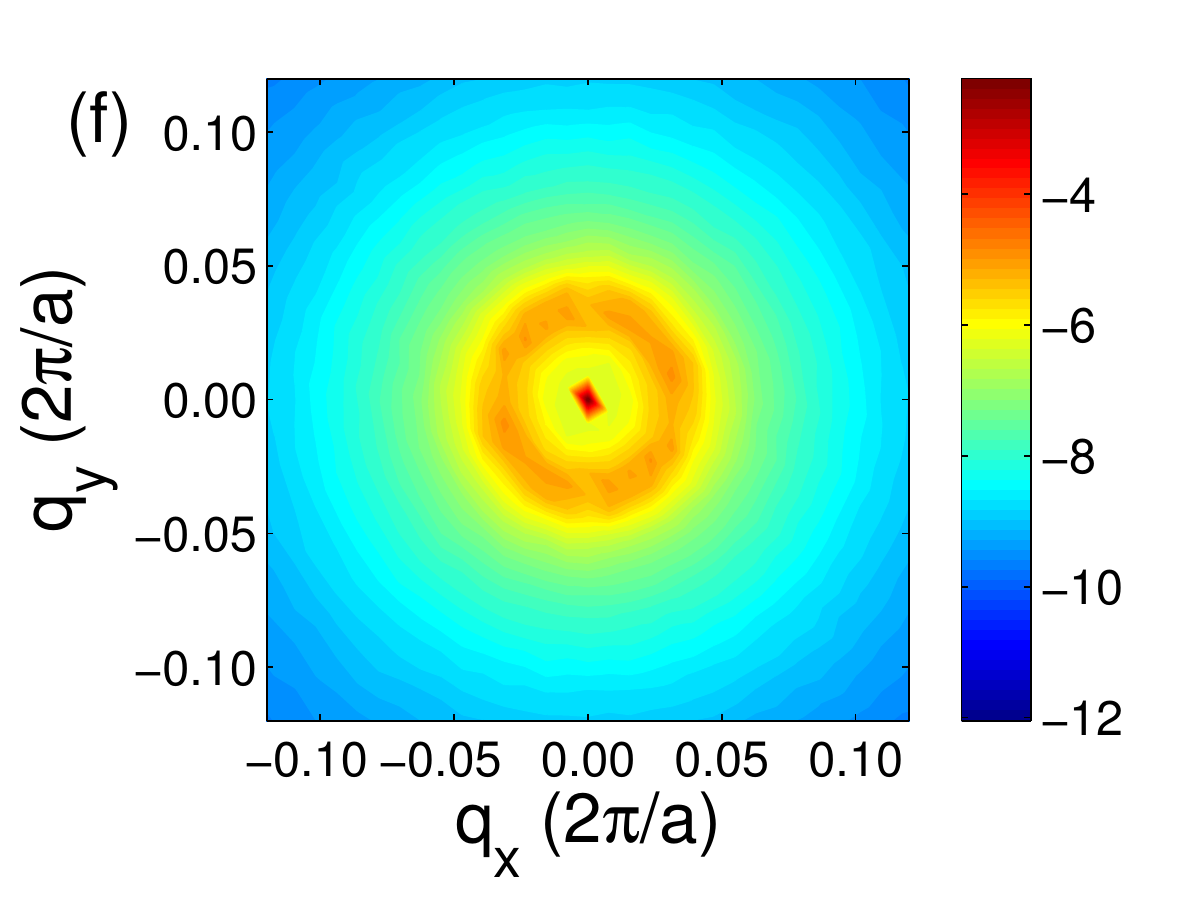}
\caption{(color online) Spin configurations of the different phases in (a)-(c) real space and (d)-(f) reciprocal space ($S\left(\boldsymbol{q}\right)$ on logarithmic scale). In the real space figures, red and blue colors denote positive and negative out-of-plane components, respectively. Shown are (a),(d) the SS ($B=0.6\,\textrm{T},T=15\,\textrm{K}$), (b),(e) the SkL ($B=2\,\textrm{T},T=15\,\textrm{K}$), and (c),(f) the FD states ($B=2\,\textrm{T},T=120\,\textrm{K}$).}
\label{fig3}
\end{figure*}

It is apparent that the number of skyrmions in the system is constant below $T_{\textrm{c}}\approx 100\,\textrm{K}$, but depends on the direction of the temperature sweep as shown in the differences between Figs.~\ref{fig2} (a) and (b). The approximately constant value of $T_{\textrm{c}}$ as a function of $B$ compares with experimental results on different systems\cite{Muhlbauer,Bauer,Bauer2,Bauer3,Wilson,Kezsmarki}. As the temperature is increased at a field value corresponding to the SkL, the topological charge starts fluctuating above $T_{\textrm{c}}$ during the MCS, while the average skyrmion number remains the same. This indicates a transition into a fluctuation-disordered (FD) state\cite{Janoschek,Bauer3,Buhrandt}, where the skyrmion lifetime is finite. The average topological charge gradually approaches zero as the paramagnetic (PM) phase or the time-reversal-invariant $B=0\,\textrm{T}$ line is approached from the direction of the FD state. It should be noted that the transition between the FP and PM phases for $B \gtrapprox 5\,\textrm{T}$ is also indicated in Figs.~\ref{fig2}(a)-(b) by an increased number of skyrmions.


Fig.~\ref{fig3} shows the non-collinear phases of the system in real and reciprocal space. The $S\left(\boldsymbol{q}\right)$ images are in good agreement with neutron scattering experiments\cite{Grigoriev,Pappas,Munzer,Kezsmarki,Muhlbauer,Adams2,Janoschek} and earlier calculations\cite{Buhrandt,Ambrose}. At finite magnetic fields, there is always a peak at $\boldsymbol{q}=\boldsymbol{0}$, reflecting the finite magnetization of the system. In the SS phase (Figs.~\ref{fig3}(a) and (d)), this is accompanied by two peaks, corresponding to the ordering related to a single $\boldsymbol{q}$ vector. In the SkL state (Figs.~\ref{fig3}(b) and (e)), there are six extra peaks corresponding to the hexagonal lattice structure. Higher harmonics are also visible in these two phases since neither the SS nor the SkL represents a perfect sinusoidal modulation (cf. Refs.~\cite{Hasselberg,Kezsmarki,Adams2}). In the FD regime (Figs.~\ref{fig3}(c) and (f)), the central peak of $S\left(\boldsymbol{q}\right)$ is surrounded by a circle, while the real-space image shows skyrmion-like spin configurations with topological charge $Q=-1$ and similar radii. Within the applied Heisenberg model, Fig.~\ref{fig3}(f) indicates that the skyrmions still have a typical equilibrium size, but they no longer form a lattice since they may be created or destroyed due to the strong thermal fluctuations above $T_{\textrm{c}}$ (see also Appendix~\ref{secS2}).


The $B-T$ phase diagram of the system is shown in Fig.~\ref{fig2}(c). The boundaries of the FD phase at higher temperatures and magnetic fields were identified from the inflection points of the $\chi\left(T\right)$ and $\chi\left(B\right)$ curves, respectively\cite{Bauer2}. We note that the FD region in this system is significantly wider ($\approx150\,\textrm{K}$) than in MnSi ($1-2\,\textrm{K}$). Note that a similarly wide transition region has been identified in MnGe at $B=0\,\textrm{T}$\cite{Deutsch,Altynbaev}, though the given interpretation was different. The transition between the FP and the PM phase was obtained similarly from $\chi\left(T\right)$, corresponding to the region with the increased number of skyrmions at large magnetic field in Figs.~\ref{fig2}(a)-(b). The lower boundary of the FD phase $T_{\textrm{c}}$ was determined by the fact that the topological charge was no longer constant during the simulation above that temperature.

As can be inferred from Figs.~\ref{fig2}(a)-(b), the range of magnetic field with high skyrmion number is wider in the simulations performed at decreasing temperature (Fig.~\ref{fig2}(b), $1\,\textrm{T}\lessapprox B \lessapprox 4\,\textrm{T}$) than in the ones with increasing temperature (Fig.~\ref{fig2}(a), $1.4\,\textrm{T}\lessapprox B \lessapprox 3\,\textrm{T}$). This indicates that in the regions SS$\rightarrow$SkL and FP$\rightarrow$SkL in Fig.~\ref{fig2}(c), for a fixed value of external field, the SkL becomes more favorable at higher temperature compared to either the SS or the FP state. This observation is in agreement with a simple Clausius--Clapeyron model of the phase boundary\cite{Bauer3},
\begin{eqnarray}
\frac{\textrm{d}T}{\textrm{d}B}=-\frac{\Delta M}{\Delta S}.
\end{eqnarray}

Since the SkL has higher entropy than both the SS and the FP states (stabilizing it in MnSi at finite temperatures\cite{Muhlbauer,Bauer3}), the slope is determined by
the magnetization, which increases by going from the SS through the SkL to the FP state. This leads to a positive slope of the transition line between the FP and the SkL states and a negative slope between the SkL and the SS states, in agreement with our previous assumption. This largely agrees with the experiments and simulations performed on GaV$_{4}$S$_{8}$ in Ref.~\cite{Kezsmarki}, but both transition lines have a negative slope for Fe$_{0.5}$Co$_{0.5}$Si thin layers as reported in Ref.~\cite{Yu}. Unfortunately, we cannot determine these lines of first-order phase transitions from MCS due to the metastability of all considered ordered states: the obtained phase will always strongly depend on the path taken in the $B-T$ space\cite{Buhrandt}.
We also experienced this effect when the sweeps were executed by changing the magnetic field at a given temperature. Note that this uncertainty due to metastability also appears under experimental conditions during field-cooling\cite{Bauer,Oike,Romming,Munzer,Milde}.

\section{Skyrmion lifetime in the fluctuation-disordered state}

\begin{figure}
\includegraphics[width=0.95\columnwidth]{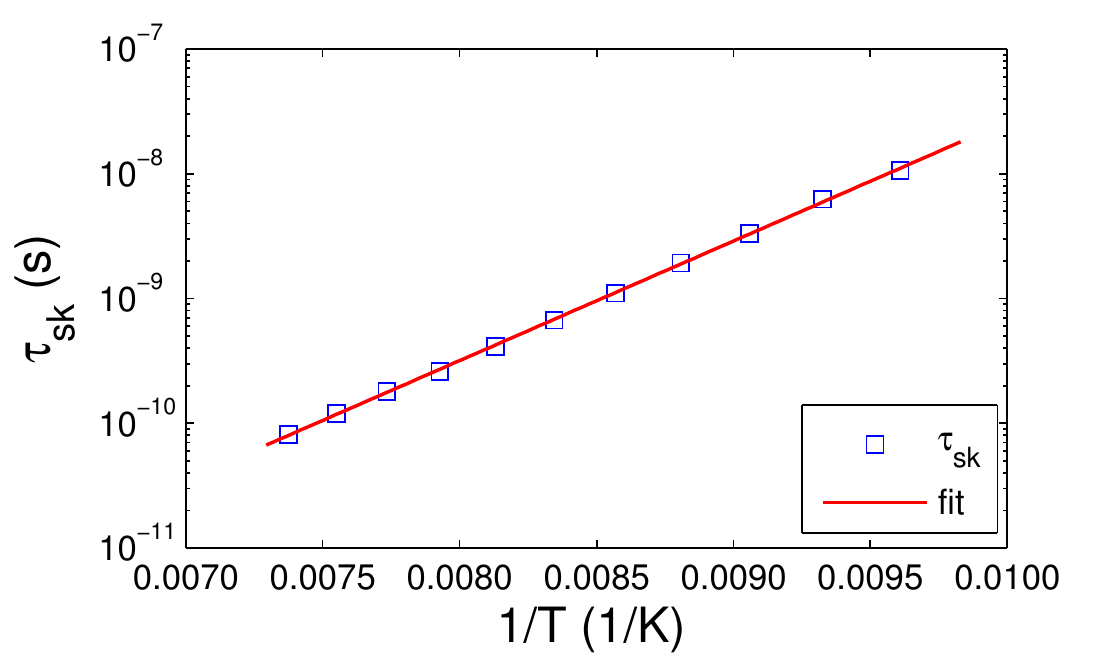}
\caption{(color online) Calculated skyrmion lifetime $\tau_{\textrm{sk}}$ as a function of temperature implying a good agreement with the Arrhenius law, Eq.~(\ref{eqn9}). The simulations were performed at $B=2\,\textrm{T}$ with a Gilbert damping constant $\alpha=0.05$. The fitted value of the energy barrier is $\frac{1}{k_{\textrm{B}}}\Delta E\approx 2200\,\textrm{K}$.}
\label{fig6}
\end{figure}

Since it was found that the skyrmion number is fluctuating in the system above $T_{\textrm{c}}$, we performed SDS to calculate the lifetime of the skyrmions $\tau_{\textrm{sk}}$ as a function of temperature at $B=2\,\textrm{T}$ -- see Appendix~\ref{secS5} for the calculation method. As shown in Fig.~\ref{fig6}, $\tau_{\textrm{sk}}$ follows the Arrhenius law,
\begin{eqnarray}
\tau_{\textrm{sk}}=\tau_{0}\textrm{e}^{\frac{\Delta E}{k_{\textrm{B}}T}},\label{eqn9}
\end{eqnarray}
in agreement with earlier works\cite{Yin,Oike,Schutte}. This is expected as skyrmion creation is a nucleation process for which Eq.~(\ref{eqn9}) usually holds. In our model, the skyrmions in the transition region are metastable in the sense that they have a finite lifetime because of their finite size, but the system in this region contains a finite number of skyrmions in the equilibrium, meaning that they are not necessarily energetically unfavorable. Actually, the system forms a SkL ground state at this value of the external field. This must be contrasted with Refs.~\cite{Yin,Oike,Schutte}, where the presence of skyrmions is energetically unfavorable with respect to the SS state or the FP state. In our calculations, we obtained $\frac{1}{k_{\textrm{B}}}\Delta E\approx 2200\,\textrm{K}$. This can be understood by the following simple argument:
The isotropic exchange interaction between the nearest neighbors, which is by far the strongest interaction between the spins, is $\frac{1}{k_{\textrm{B}}}J_{1}\approx 400\,\textrm{K}$\cite{Simon}. The skyrmion is unwound if the downwards pointing spin in the middle of the skyrmion is rotated in the direction parallel to the external field, during which rotation it loses the exchange energy stabilizing it with respect to its nearest neighbors, which are almost parallel to it in the initial state.
This energy difference equals approximately $\Delta E\approx6J_{1}$, since there are six nearest neighbors in the triangular lattice. We note that in the simulations performed on a square lattice in Ref.~\cite{Yin} $\Delta E\approx 2J_{1}$ was approximated by taking into account only a single triangle of spins in the energetical and topological considerations, and good agreement with the corresponding simulation results was obtained. Although in bulk systems skyrmions correspond to lines along the direction of the external field, they are unwound as a consequence of the appearance of local defects (monopoles\cite{Milde}), with an energy barrier of $\Delta E\approx 5.8J_{1}$ on a simple cubic lattice (also six nearest neighbors) reported in Ref.~\cite{Schutte}. Experimentally, Ref.~\cite{Oike} reports $\frac{1}{k_{\textrm{B}}}\Delta E\approx 2000\,\textrm{K}$ for MnSi, where the exchange interactions are probably weaker than in Pd/Fe/Ir$(111)$, indicated by the lower critical temperature $T_{\textrm{c}}\approx 29\,\textrm{K}$.

In our simulations, we obtained $T_{\textrm{c}}\approx 100\,\textrm{K}$, which meant that the skyrmion number does not fluctuate under the timescale accessible with SDS, $t\approx100\,\textrm{ns}$ (see Fig.~\ref{fig6}) or for similar equivalent time scales available in MCS. If Eq.~(\ref{eqn9}) holds over a much wider range of temperature, it can be extrapolated that the skyrmion lifetime reaches the $1\,\textrm{s}$ scale around $T'_{\textrm{c}}\approx50\,\textrm{K}$. Fluctuations on this timescale are accessible to experimental methods as has been demonstrated in Ref.~\cite{Oike}, and the experimentally determined critical temperature may be closer to this value. The generation of skyrmions due to current injection was examined for Pd/Fe/Ir$(111)$  in Ref.~\cite{Romming}, where it was concluded that the skyrmions are stable against thermal excitations at $T=4.2\,\textrm{K}$, in agreement with our simulations.
As a comparison, the ordering temperature reported for the similar system Fe/Ir$(111)$ in Ref.~\cite{Sonntag} is $T_{\textrm{c}}\approx 28\,\textrm{K}$. Although this is lower even than the extrapolated value of $T'_{\textrm{c}}\approx50\,\textrm{K}$, it is known from \textit{ab initio} calculations\cite{Dupe,Simon} that the exchange interactions determining the ordering temperature are significantly weaker in the absence of the Pd overlayer.


\section{Conclusion}

We expect that the $B-T$ phase diagram of PdFe bilayer on Ir$(111)$ surface determined in this paper is qualitatively similar for ultrathin film systems within the same symmetry class, as long as the ground state of the system at $B=0\,\textrm{T}$ remains the SS state. This similarity was already demonstrated for bulk systems with different compositions in Refs.~\cite{Bauer,Tokunaga}. Since the interaction parameters can be tuned by changing the composition as was demonstrated in Ref.~\cite{Dupe}, it should be possible to change the skyrmion lifetime within the FD state, being of crucial importance in technological applications.


We note that after the submission of our work, the paper Ref.~\cite{Hagemeister} was published, also concerning the finite lifetime of skyrmions in the Pd/Fe/Ir$(111)$ system. This study focuses on the FP$\rightarrow$SkL region of the phase diagram in Fig.~\ref{fig2}(c) where individual skyrmions are excited in the homogeneous FP state, while we determined the skyrmion lifetime at the transition from the SkL to the FD state where there is a large number of skyrmions in the system, necessitating different simulation methods (see also Appendix~\ref{secS5}).


\begin{acknowledgments}
The authors thank E. Y. Vedmedenko and R. Wiesendanger for enlightening discussions. Financial support for this work was provided by the Hungarian Scientific Research Fund under project Nos.~K115575 and K115632. K. P. acknowledges support from the Hungarian E\"{o}tv\"{o}s Fellowship.
\end{acknowledgments}

\appendix

\section{\textit{Ab initio} calculations\label{secS1}}




The screened Korringa--Kohn--Rostoker method\cite{Szunyogh,Zeller} was applied to calculate the electronic structure of the system self-consistently, first for the Ir bulk and then for a layered geometry of nine Ir, one Fe, one Pd and four vacuum (empty cell) layers on top of the $(111)$ surface of the bulk in fcc growth. The calculations used the local spin density approximation with the potential parametrization given in Ref.~\cite{Vosko}, and the atomic sphere approximation. We only considered a geometry optimized by VASP calculations\cite{Kresse,Kresse2,Hafner}, corresponding to $5\%$ relaxation of the Fe layer with respect to the top Ir layer in Ref.~\cite{Simon}. In good agreement with the values recently reported in Ref.~\cite{Crum}, we obtained a spin magnetic moment of $2.95\,\mu_{\textrm{B}}$ for the Fe atoms. The spin-cluster expansion combined with the relativistic disordered local moment scheme\cite{Szunyogh2} was used to calculate the coupling coefficients in the spin model, Eq.~(\ref{eqn1}). The induced moments in the Pd and Ir layers disappear in the paramagnetic phase modeled by the relativistic disordered local moments scheme, therefore, in the spin model we restricted ourselves to the Fe moments. The coupling coefficients were calculated for $240$ neighbors within the circle of radius $8a$, where $a=2.71\,\textrm{\AA}$ is the lattice constant of the triangular lattice on the Ir$(111)$ surface. For a thorough discussion of the calculated couplings see Ref.~\cite{Simon}.

\section{Simulation methods\label{secS2}}

We examined the equilibrium properties of the PdFe bilayer at finite temperatures and external magnetic fields in terms of classical Monte Carlo simulations using the Metropolis algorithm. We also performed spin dynamics simulations based on the numerical solution of the stochastic Landau--Lifshitz--Gilbert equation\cite{Landau,Gilbert,Brown,Kubo}
\begin{eqnarray}
\frac{\textrm{d}\boldsymbol{S}_{i}}{\textrm{d}t}&=&-\gamma' \boldsymbol{S}_{i} \times \left(\boldsymbol{B}_{i}^{\textrm{eff}}+\boldsymbol{B}_{i}^{\textrm{th}}\right) \nonumber
\\
&&- \gamma' \alpha\boldsymbol{S}_{i} \times\left[\boldsymbol{S}_{i} \times \left(\boldsymbol{B}_{i}^{\textrm{eff}}+\boldsymbol{B}_{i}^{\textrm{th}}\right) \right].\label{eqnS1}
\end{eqnarray}

Here $\alpha$ denotes the Gilbert damping parameter, $\gamma=\frac{ge}{2m}$ the gyromagnetic ratio ($g,e,m$ are the electronic spin $g$-factor, charge, and mass, respectively), $\gamma'=\frac{\gamma}{1+\alpha^{2}}$, $\boldsymbol{B}_{i}^{\textrm{eff}}=-\frac{1}{m}\frac{\partial H}{\partial \boldsymbol{S}_{i}}$ the effective magnetic field at site $i$ and $\boldsymbol{B}_{i}^{\textrm{th}}(t)=\sqrt{\frac{2\alpha k_{\textrm{B}}T}{m\gamma}}\circ\boldsymbol{\eta}_{i}(t)$ the thermal noise. The $\circ$ symbol denotes Stratonovich stochastic calculus used during the integration of the equation of motion\cite{Garcia-Palacios}. The integration was performed using the so-called semi-implicit B method discussed in Ref.~\cite{Mentink}, which conserves the length of the spins.

Generally, the Monte Carlo simulations require less computation time for thermalization and calculation of averages with a given precision than the spin dynamics simulations. Spin dynamics simulations can be used to find the ground state at zero temperature where the Boltzmann factor in the Monte Carlo method becomes singular. Even more importantly, the time step in spin dynamics simulations is trivially scaled to real time once the constants in Eq.~(\ref{eqnS1}) are known, making it possible to evaluate the real-time-dependence of quantities such as the topological charge $Q$.

The Monte Carlo simulations were performed at a fixed value of the external magnetic field, while the temperature was increased or decreased (Figs.~\ref{fig2}(a)-(b)). At every point in the $B-T$ space, the system was thermalized for $2\cdot10^{5}$ Monte Carlo steps, and the averages were obtained from $10^{6}$ Monte Carlo steps. For the calculation of the skyrmion lifetime in Appendix~\ref{secS5}, we performed spin dynamics simulations of length $10^{7}-10^{8}t_{0}$ at every temperature, with the time unit $t_{0}=\frac{\hbar}{2\textrm{mRyd}}=2.42\cdot10^{-14}\,\textrm{s}$, and the time step $\Delta t=0.01t_{0}$. The simulation length was sufficiently long to obtain multiple skyrmion creation and annihilation events at every considered temperature during a single run. The damping parameter was $\alpha=0.05$. The calculations were performed for an $N=128\times128$ lattice with periodic boundary conditions.

\begin{figure*}
\includegraphics[width=\columnwidth]{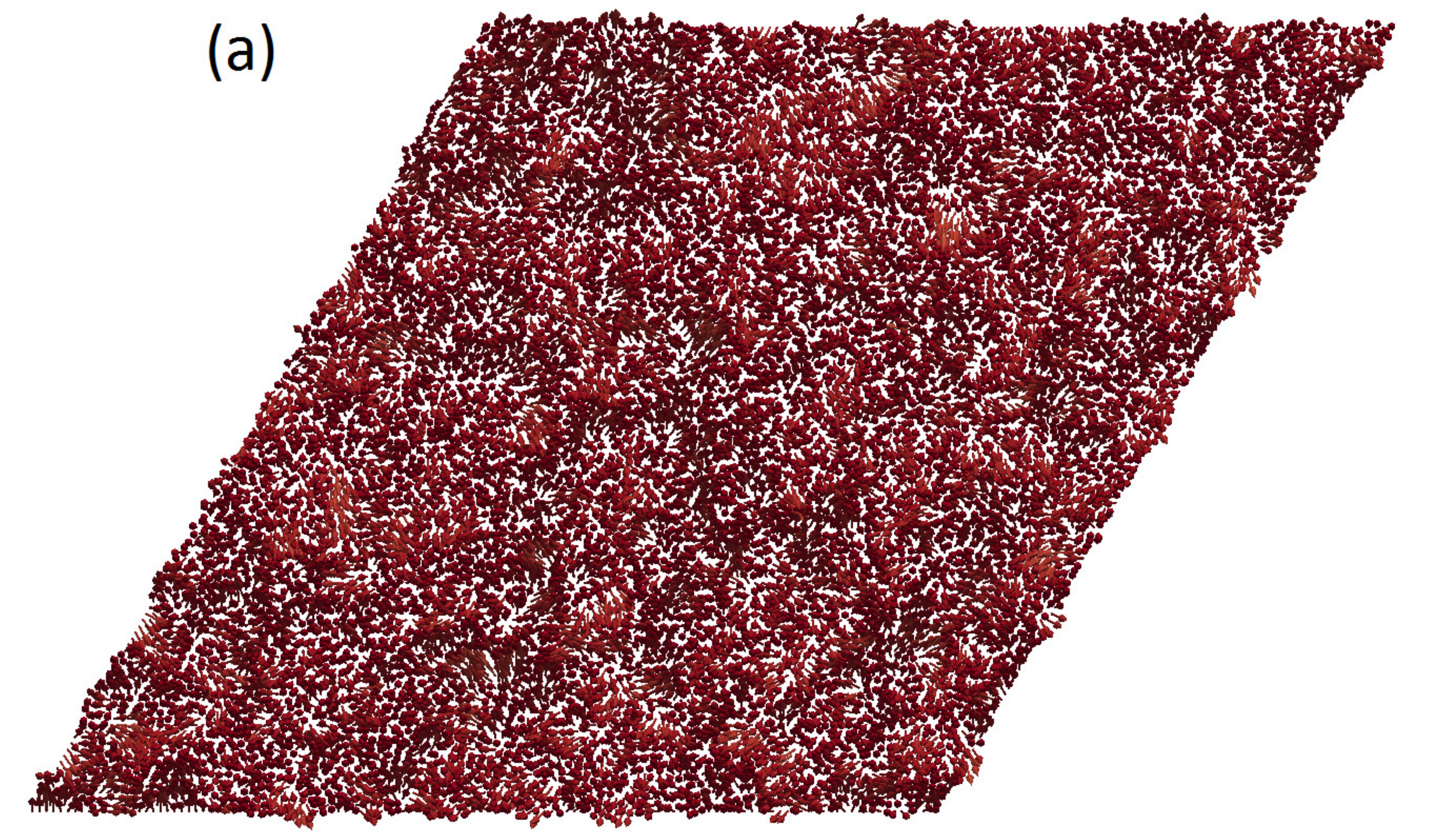}
\includegraphics[width=\columnwidth]{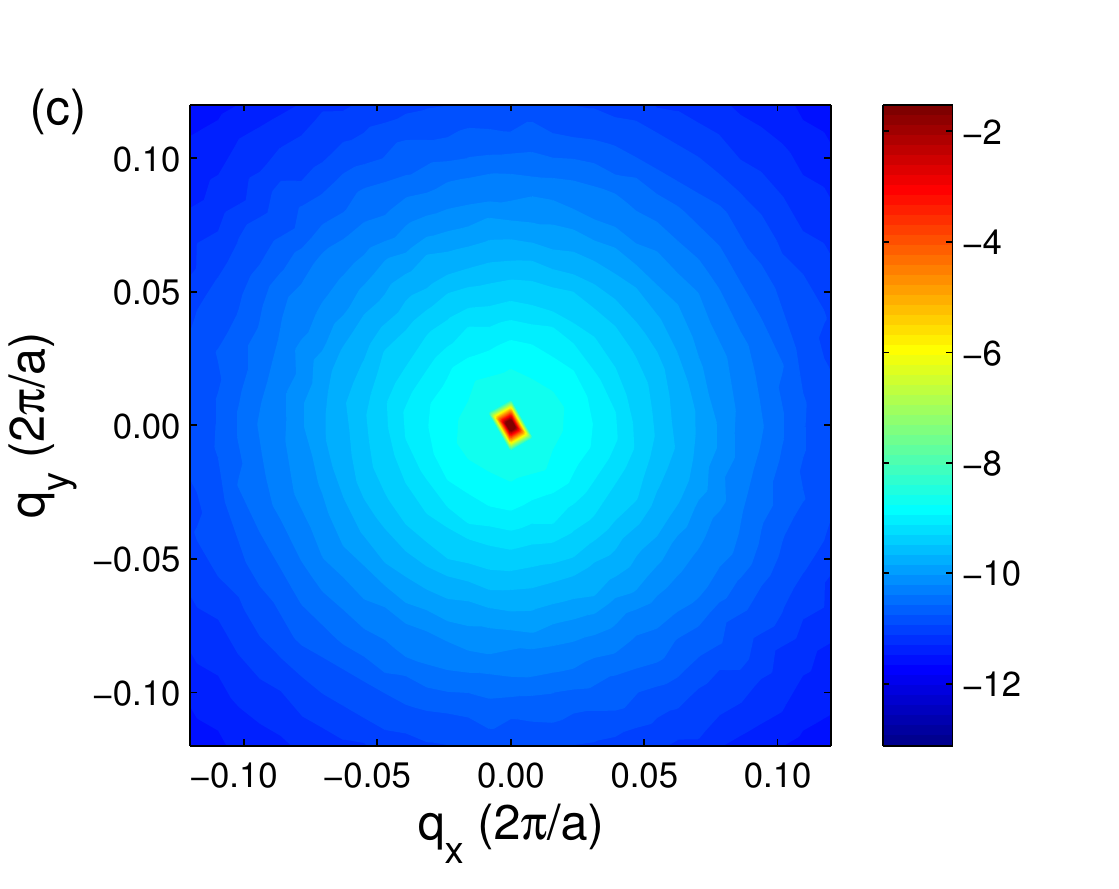}
\includegraphics[width=\columnwidth]{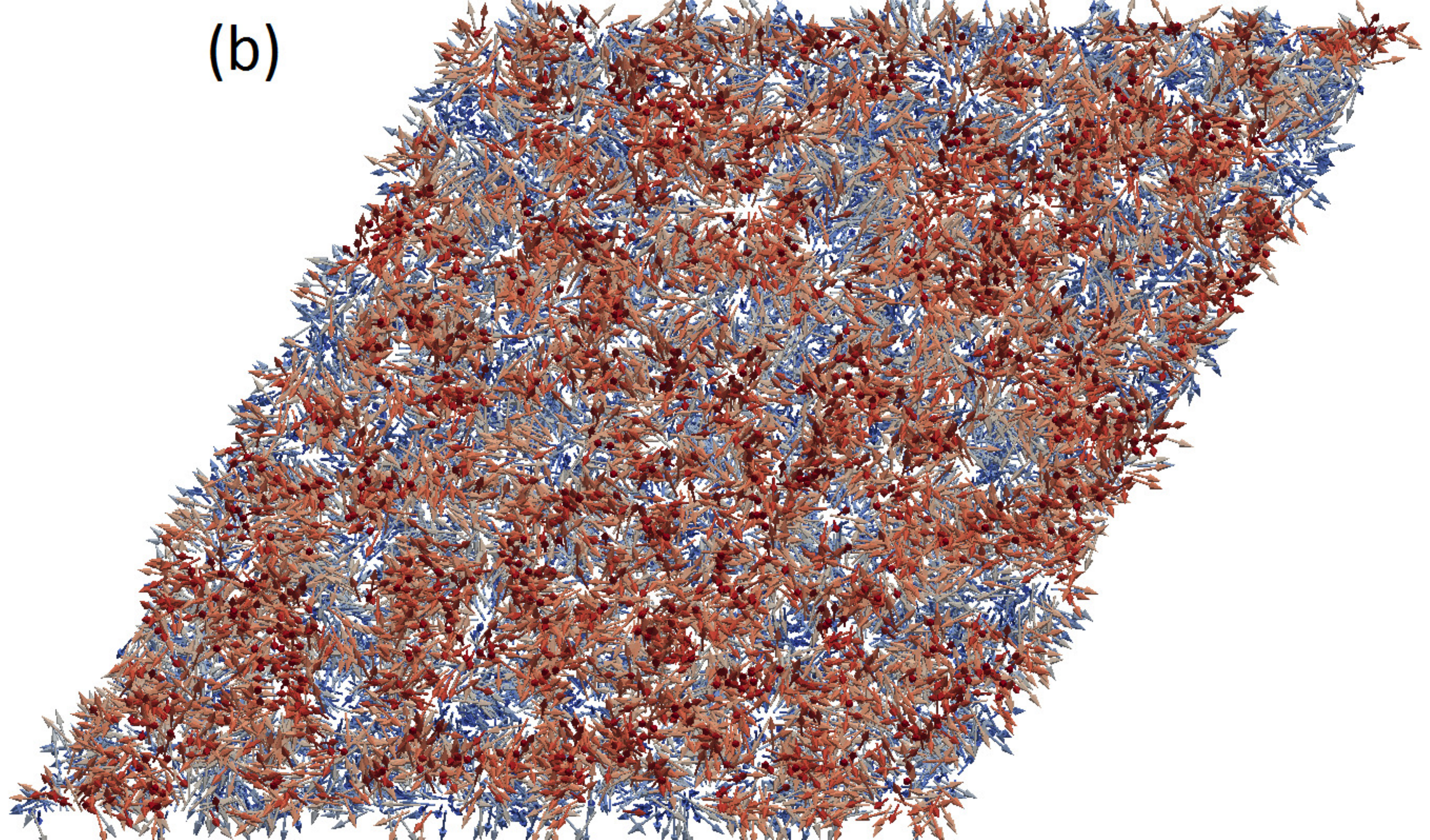}
\includegraphics[width=\columnwidth]{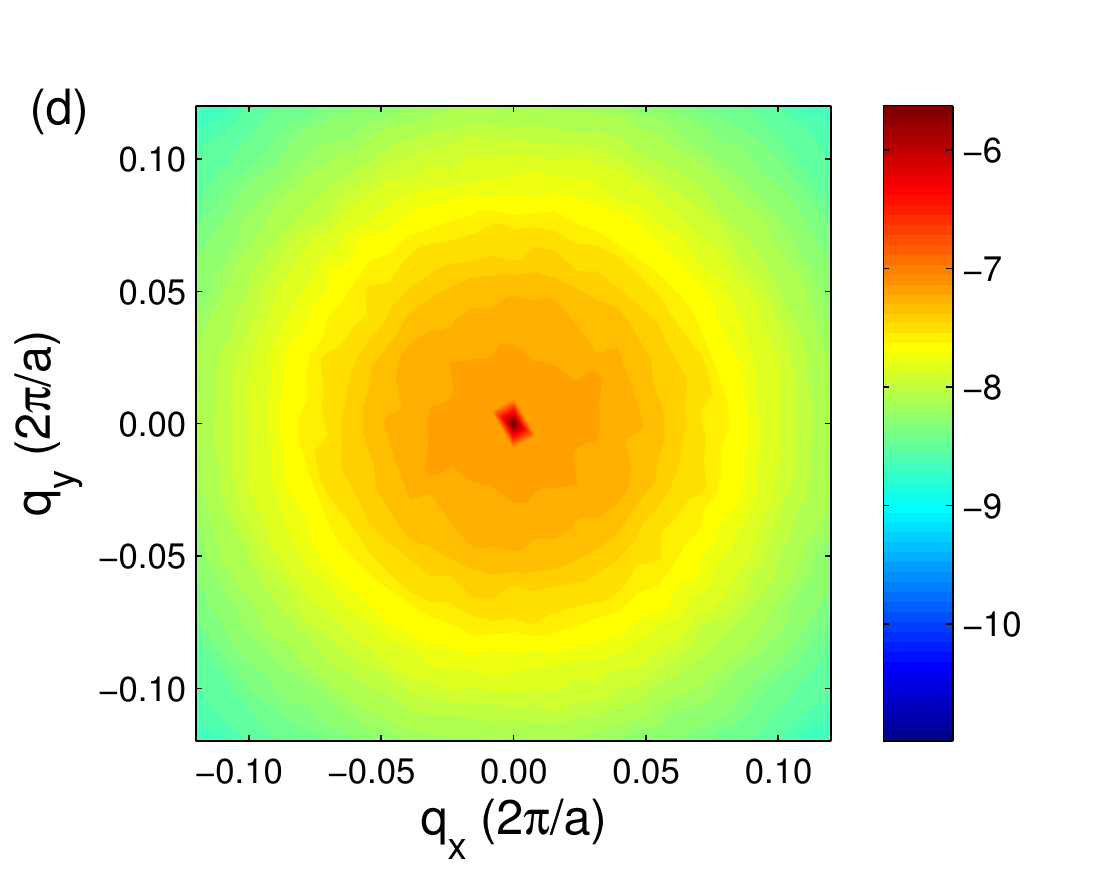}
\caption{(color online) Spin configurations of the (a),(c) field-polarized ($B=3.2\,\textrm{T},T=15\,\textrm{K}$) and the (b),(d) paramagnetic ($B=2\,\textrm{T},T=310\,\textrm{K}$) phases in real space and reciprocal space ($S\left(\boldsymbol{q}\right)$ on logarithmic scale).}
\label{figS1}
\end{figure*}

The magnetic phases can be characterized by the 
static structure factor,
\begin{eqnarray}
S\left(\boldsymbol{q}\right)=\left<\sum_{\alpha}\left|\frac{1}{N}\sum_{i}\textrm{e}^{\textrm{i}\boldsymbol{q}\boldsymbol{R}_{i}}S_{i}^{\alpha}\right|^{2}\right>,\label{eqnS2}
\end{eqnarray}
corresponding to the intensity profile in elastic neutron scattering experiments\cite{Baruchel}. $S\left(\boldsymbol{q}\right)$ has peaks corresponding to the reciprocal lattice vectors of the magnetic order of the system. Since the characteristic length scale of the considered non-collinear states (spin spiral, skyrmion lattice) is significantly larger than the lattice constant $a$, these peaks appear inside the first Brillouin zone of the atomic lattice, close to the $\Gamma$ point. Eq. (\ref{eqnS2}) indicates the $S\left(\boldsymbol{q}\right)=S\left(-\boldsymbol{q}\right)$ symmetry of the structure factor; therefore, the peaks at $\boldsymbol{q}\neq\boldsymbol{0}$ always appear in pairs, even when the spin structure can be described by a single wave vector as in the case of the spin spiral phase. To complement  Fig.~\ref{fig3} related to the non-collinear phases, in Fig.~\ref{figS1} we present the field-polarized and paramagnetic phases in real and reciprocal space. The field-polarized state Fig.~\ref{figS1}(c) shows no specific feature beyond the peak at $\boldsymbol{q}=\boldsymbol{0}$, while in the paramagnetic phase Fig.~\ref{figS1}(d) this is broadened to a Lorentzian curve.

We note that the neutron scattering profile in Fig.~\ref{fig3}(f) may also indicate the presence of stable skyrmions with infinite lifetime which show no long-range order either due to their thermal motion as in a skyrmion liquid\cite{Pappas2} or their random arrangement as in an amorphous solid\cite{Rossler}. We indeed identified the same profile below $T_{\textrm{c}}$ in some parts of the SkL phase, but this is probably due to the reduced dimensionality of the system, as an amorphous arrangement of skyrmions was also identified experimentally in real space in thin films in Refs.~\cite{Romming,Yu}. We did not partition the skyrmion lattice phase into further subphases since the mentioned neutron scattering profile is not accompanied by the variation of $Q$ during the simulation, unlike in the fluctuation-disordered state. We also point out that in Ref.~\cite{Rossler}, the amorphous skyrmion state at $B=0\,\textrm{T}$ was explained within a model that accounts for longitudinal spin fluctuations close to the critical temperature; in the Heisenberg model with fixed moment length, we only identified the spin spiral, fluctuation-disordered and paramagnetic phases for zero external field.

\begin{figure}
\includegraphics[width=\columnwidth]{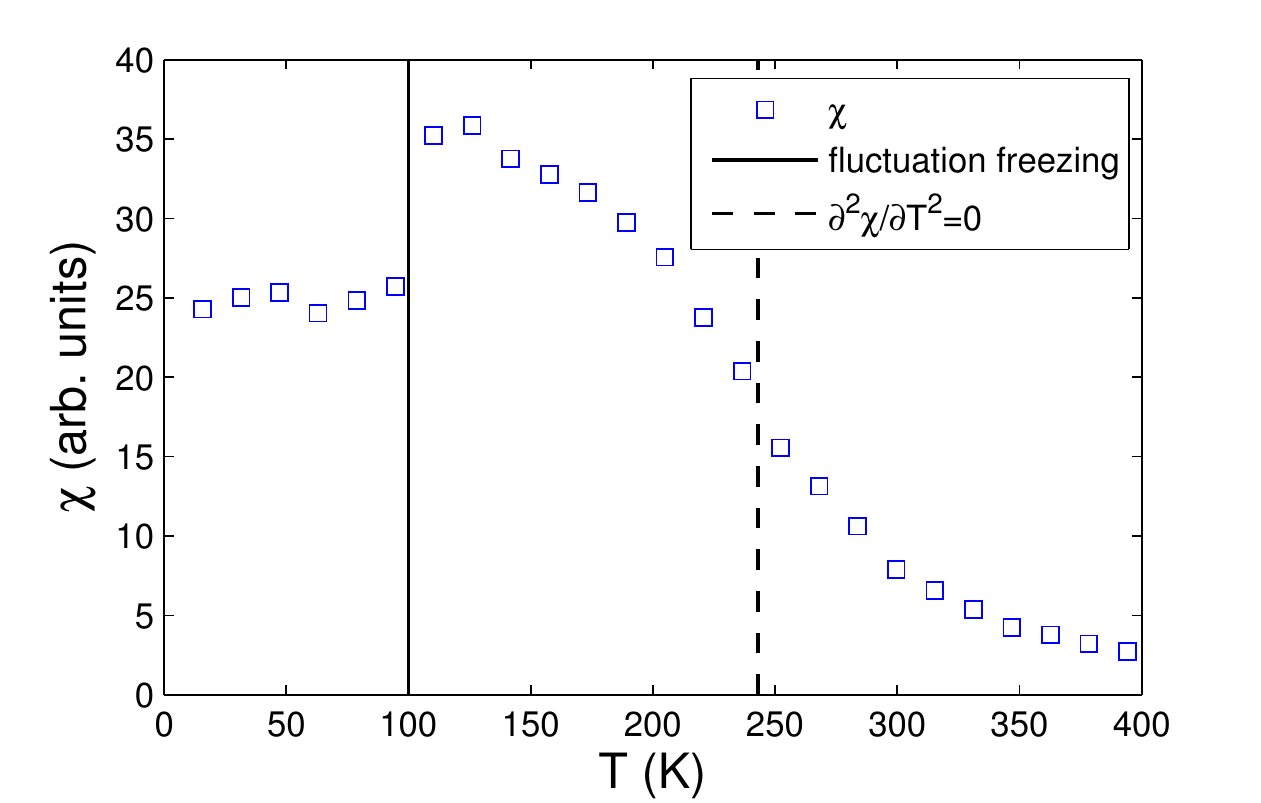}
\caption{(color online) Static magnetic susceptibility $\chi$ as a function of temperature for $B=0\,\textrm{T}$. Horizontal lines denote the phase boundaries of the fluctuation-disordered phase, corresponding to a singularity (solid line) and an inflection point (staggered line) in $\chi$.}
\label{figS2}
\end{figure}

The static magnetic susceptibility
\begin{equation}
\chi=\frac{1}{k_{\textrm{B}}T}\left(\left<\boldsymbol{M}^{2}\right>-\left<\boldsymbol{M}\right>^{2}\right),
\end{equation}
with
\begin{equation}
\boldsymbol{M}=\frac{1}{N}\sum_{i}\boldsymbol{S}_{i} ,
\end{equation}
is appropriate for finding the transition points between the different phases\cite{Bauer2}. At a first-order transition it generally exhibits a finite jump or fall, while second-order transition points correspond to inflection points in $\chi$ as a function of temperature. One example is shown in Fig.~\ref{figS2} at $B=0\,\textrm{T}$, showing similarity to the curve calculated from the Brazovski\u{i} model in Ref.~\cite{Janoschek}. The inflection points were determined by fitting a polynomial on the $\chi\left(T\right)$ and $\chi\left(B\right)$ curves. Although the calculated inflection points give a good first approximation of the phase transition lines as shown in this paper, the detailed comparison of several different thermodynamical quantities should improve the accuracy of phase transition lines as was demonstrated in Refs.~\cite{Bauer2,Demishev} for MnSi.

\section{Topological charge\label{secS3}}

The skyrmion number is connected to the topological charge of the spin system. In field theory, this is given by\cite{Belavin}
\begin{eqnarray}
Q=\frac{1}{4\pi}\int\boldsymbol{S}\left(\partial_{x}\boldsymbol{S}\times\partial_{y}\boldsymbol{S}\right)\textrm{d}x\textrm{d}y,\label{eqnS5}
\end{eqnarray}
where $\boldsymbol{S}$ is a vector field normalized to $1$. If one introduces the usual polar and azimuthal angles
\begin{eqnarray}
\boldsymbol{S}\left(x,y\right)=\left[\begin{array}{c}\sin\vartheta\left(x,y\right)\cos\varphi\left(x,y\right) \\ \sin\vartheta\left(x,y\right)\sin\varphi\left(x,y\right) \\ \cos\vartheta\left(x,y\right)\end{array}\right],
\end{eqnarray}
Eq.~(\ref{eqnS5}) transforms into
\begin{eqnarray}
Q=\frac{1}{4\pi}\int\sin\vartheta\left(x,y\right)\frac{\partial\left(\vartheta,\varphi\right)}{\partial\left(x,y\right)}\textrm{d}x\textrm{d}y,\label{eqnS7}
\end{eqnarray}
where $\frac{\partial\left(\vartheta,\varphi\right)}{\partial\left(x,y\right)}$ is the signed Jacobian determinant. This demonstrates that $Q$ counts how many times $\boldsymbol{S}$ winds around the unit sphere.

The correct way of generalizing this quantity to the lattice model discussed in this paper is given in Ref.~\cite{Berg} (see also Ref.~\cite{Yin}). The lattice must be partitioned into nearest-neighbor triangles of spins, and one must sum up the signed areas of the spherical triangles defined by the $\boldsymbol{S}_{i}$ vectors. If the vectors are denoted by $\boldsymbol{S}_{1},\boldsymbol{S}_{2},\boldsymbol{S}_{3}$ following a counterclockwise rotation on the lattice, the sign of the area is the same as the sign of the product $\boldsymbol{S}_{1}\left(\boldsymbol{S}_{2}\times\boldsymbol{S}_{3}\right)$. If the calculations are performed for a lattice with periodic boundary conditions, one will always return to the same point in the lattice during the calculation, and the calculated spherical area will always be an integer multiple $Q$ of the area of the whole sphere, $4\pi$. By calculating the skyrmion number this way, the fluctuations in the topological charge at finite temperature will always signal actual topological changes, rather than discretization errors, making it possible to calculate the skyrmion lifetime.

\begin{figure}
\includegraphics[width=0.3\columnwidth]{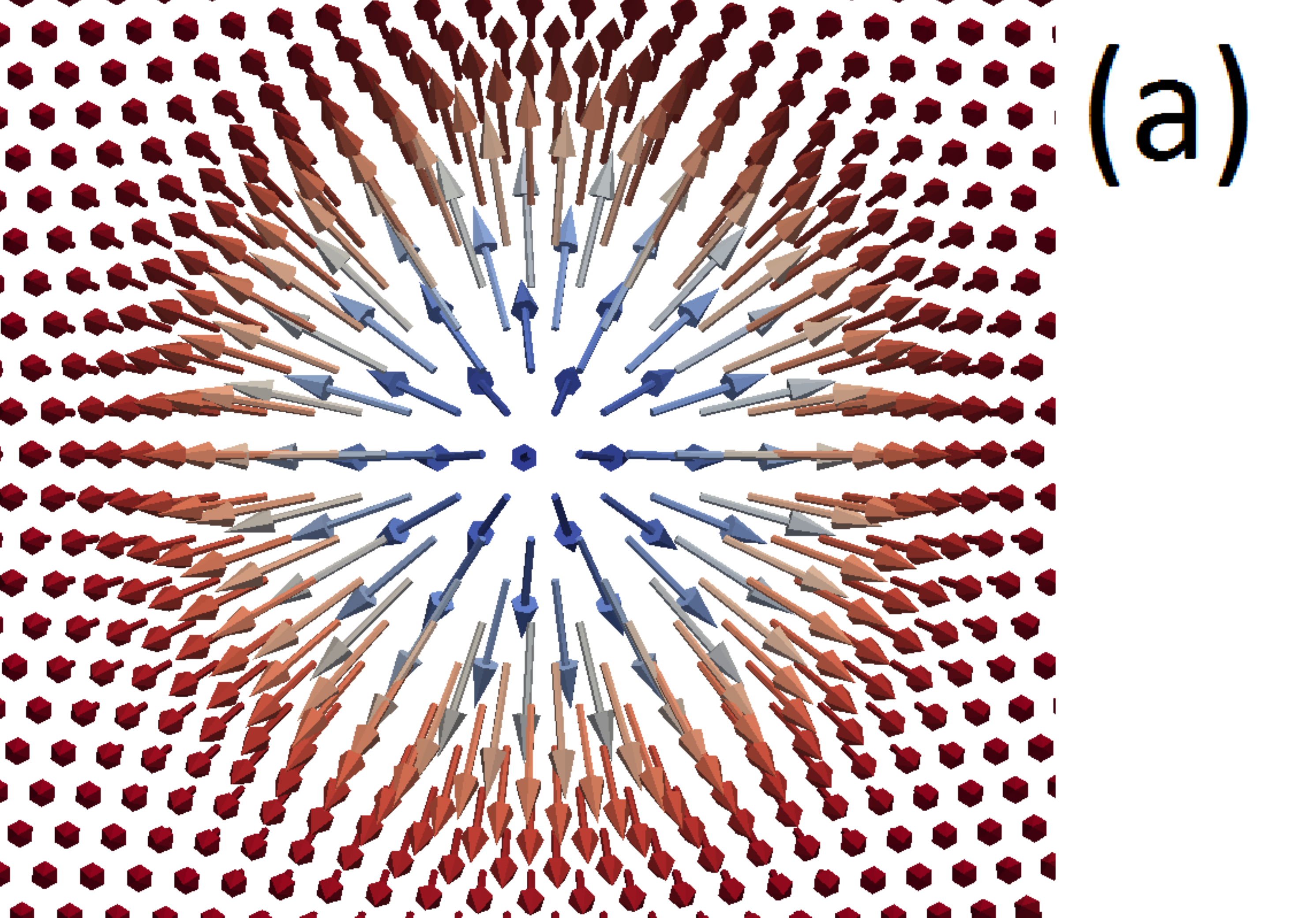}
\includegraphics[width=0.3\columnwidth]{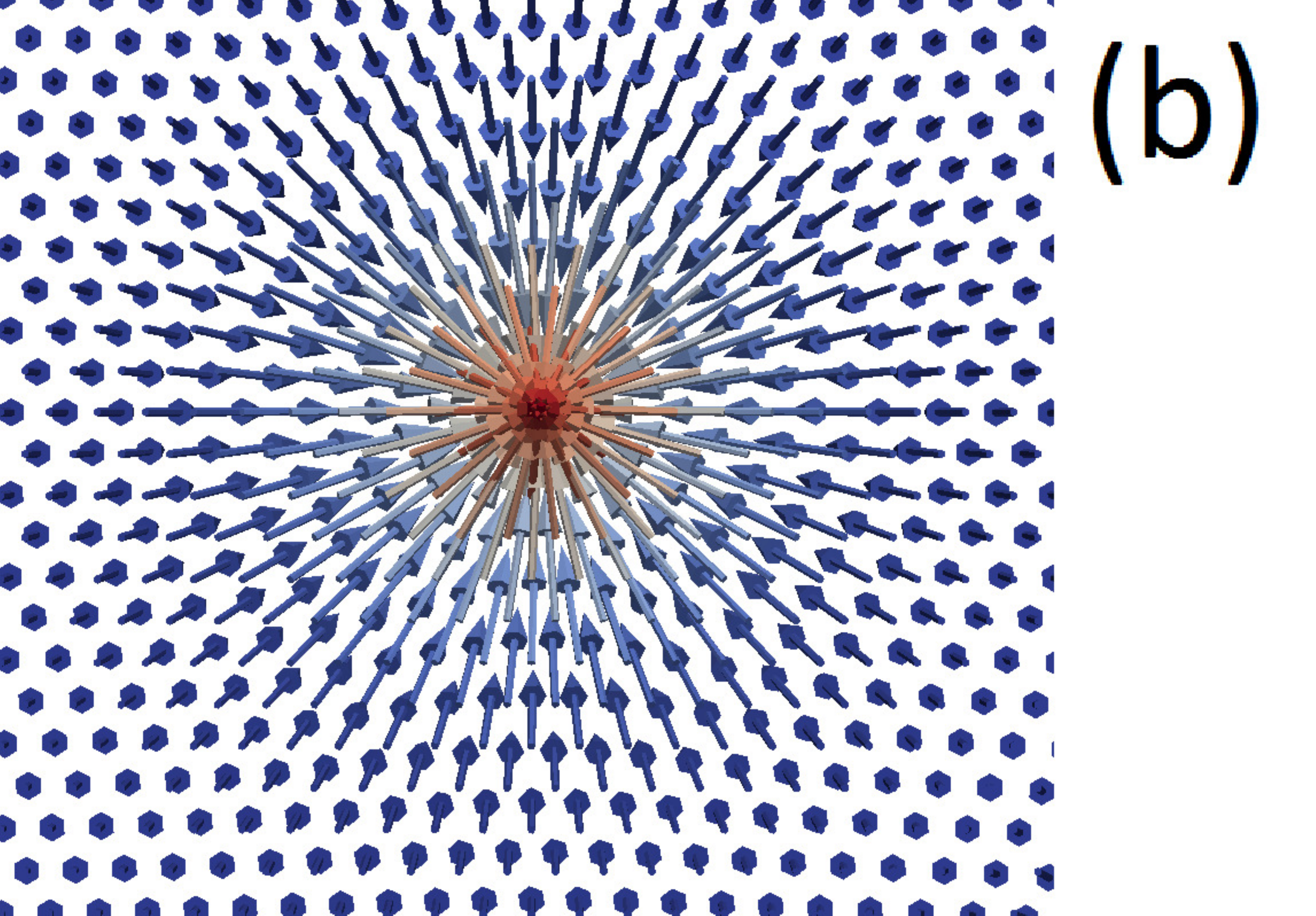}
\includegraphics[width=0.3\columnwidth]{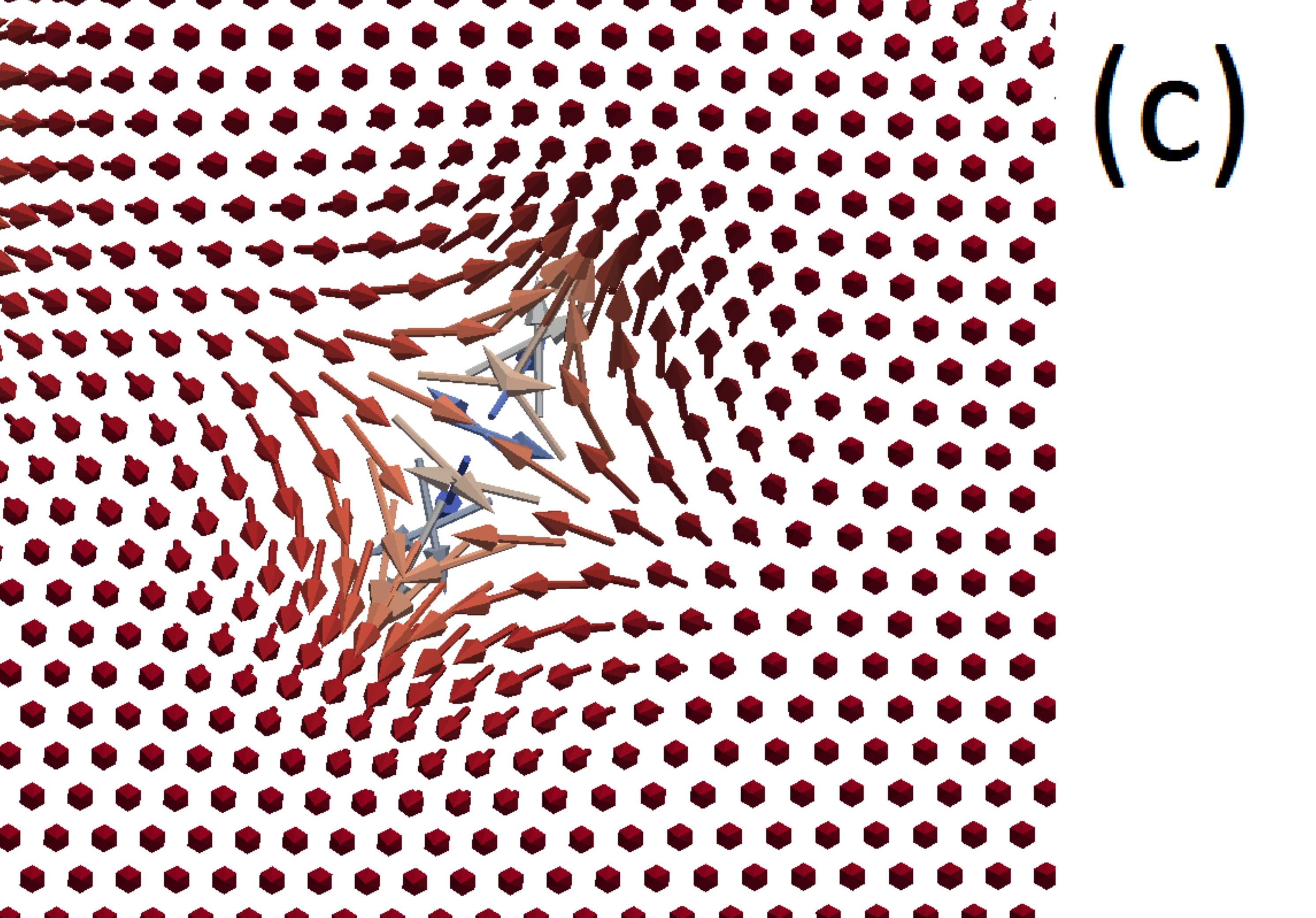}
\caption{(color online) Localized spin configurations with different topological charges: (a) downwards pointing skyrmion ($Q=-1$), (b) upwards pointing skyrmion ($Q=1$) and (c) downwards pointing antiskyrmion ($Q=1$). The direction refers to the fact that the spin at the center of the skyrmion is pointing towards (down) or outwards from (up) the Ir surface. For $\boldsymbol{B}$ pointing upwards, only downwards pointing skyrmions are energetically favorable.}
\label{figS3}
\end{figure}

The connection between the topological charge $Q$ and the skyrmion number is demonstrated in Fig.~\ref{figS3}. Firstly, Eq.~(\ref{eqnS5}) changes sign under time reversal, therefore the upwards and downwards pointing skyrmions have opposite charge. However, they also have an opposite finite magnetization perpendicular to the plane; therefore, the external field $B$ will break the energy degeneracy between them, leading to the disappearance of the upwards pointing skyrmions if the field is pointing upwards, that is outwards from the Ir surface for which direction of $\boldsymbol{B}$ the simulations were carried out. Secondly, downwards pointing antiskyrmions also have opposite charge to the skyrmions, and they have the same energy due to the isotropic Heisenberg coupling and the Zeeman term\cite{Leonov2}. The energy difference between the skyrmions and antiskyrmions is caused by the Dzyaloshinsky--Moriya interaction. For small wave vectors, the isotropic exchange and the Dzyaloshinsky--Moriya interaction energy contributions are quadratic and linear in the wave vector, respectively\cite{Bode}. Energy minimization yields that the characteristic scale of spin spirals and skyrmion lattices in reciprocal space is given by $qa\approx\frac{D}{J}$, while the energy gain per spin is approximately  $D^{2}/J$, where $D$ and $J$ are the effective Dzyaloshinsky--Moriya and isotropic exchange interactions\cite{Han}. The non-collinear spin structures only gain this energy if they follow a rotational sense specified by the Dzyaloshinsky--Moriya interaction. As shown in Figs.~\ref{figS3}(a)-(b), the rotational sense of the spins is the same for every cross section of a skyrmion, so skyrmions gain energy due to the chiral interaction. However, the rotational sense in antiskyrmions in Fig.~\ref{figS3}(c) may be either left-handed or right-handed depending on the chosen cross section, meaning that they do not gain energy from the Dzyaloshinsky--Moriya interaction. Similarly, the energy gain of the skyrmion compared to the field-polarized state is also due to the Dzyaloshinsky--Moriya interaction, but since the field-polarized state has a higher magnetization, there is a transition between the two states at approximately $mB\approx D^{2}/J$\cite{Bogdanov2}. This means that antiskyrmions are always energetically unfavorable, and upwards pointing skyrmions are also unfavorable in the case of a large enough external field ($B\gtrapprox1\,\textrm{T}$), meaning that in this region, $Q$ may be identified with the number of downwards pointing skyrmions which form the hexagonal skyrmion lattice state.

\section{The effect of periodic boundary conditions\label{secS4}}

We performed the simulations on an $N=128\times 128$ lattice with periodic boundary conditions, which determines which values of the wave vector $\boldsymbol{q}$ are allowed in the system. The wavelength of the spin spiral ($\lambda=6.6\,\textrm{nm}$) was determined from a mean-field model based on calculating the Fourier transform $\mathcal{J}\left(\boldsymbol{q}\right)$ of the coupling coefficients in Eq.~(\ref{eqn1})\cite{Simon}. Since the $(111)$ surface of the Ir fcc lattice has threefold rotational symmetry, the maximal eigenvalues of $\mathcal{J}\left(\boldsymbol{q}\right)$ are almost isotropic for small wave vectors. This degeneracy could only be broken by higher-order anisotropy terms in the Hamiltonian, which we omitted from the calculations; and the above mentioned boundary conditions, which discretize the values of $\boldsymbol{q}$ and fix the direction of the spin spiral wave vector. The lattice constant of the skyrmion lattice was $d_{c}=8.7\,\textrm{nm}$, corresponding to $32$ Ir$(111)$ lattice constants or $16$ skyrmions in the $128\times 128$ lattice. Note that $d_{c}$ is the distance between the skyrmion cores, which is different from the skyrmion core diameter $d_{0}$\cite{Butenko}.

Due to the periodic boundary conditions, $\lambda$ and $d_{c}$ could only change through topological transitions, that is an addition or removal of a $360^{\circ}$ domain wall in a spin spiral or a skyrmion. It is known from micromagnetic calculations\cite{Bogdanov2,Rossler2,Han} that both $\lambda$ and $d_{c}$ depend on the external magnetic field at zero temperature. However, this dependence is strongest close to a critical field value where the spin spiral and skyrmion lattice states become metastable compared to the field-polarized state and break into $360^{\circ}$ domain walls and isolated skyrmions, respectively. The spin spiral state was not examined close to this critical field since it becomes metastable compared to the skyrmion lattice at a significantly lower field value\cite{Han}. For the skyrmion lattice, we performed calculations for different $d_{c}$ values close to the transition into the field-polarized state ($B\approx 3\,\textrm{T}$ at $T=0\,\textrm{K}$). 
 For $B=3\,\textrm{T}$, the lattice with $d_{c}=8.7\,\textrm{nm}$ had the lowest energy out of the ones commensurate with the $N=128\times 128$ lattice; while for $B=3.2\,\textrm{T}$, even the system with a single skyrmion ($d_{c}=34.8\,\textrm{nm}$) had higher energy than the field-polarized state. This leads us to the conclusion that the divergence of $d_{c}$ happens in a very short interval of magnetic field, in agreement with Ref.~\cite{Han}. This is probably because the divergence of $d_{c}$ is a consequence of the repulsive interaction between the skyrmions, but this interaction decreases exponentially with the distance between the skyrmions outside the core areas\cite{Lin}. 

However, using free boundary conditions would not solve the problem with the relaxation of $\lambda$ and $d_{c}$ since the spins at the edges form configurations which also interact with the spin spiral and the skyrmion lattice, see e.g. Ref.~\cite{Keesman}. The topological charge is also no longer an integer for free boundaries, which complicates the determination of the skyrmion lifetime at finite temperature. Therefore, we conclude that the periodic boundary conditions are probably more appropriate for the simulation of a homogeneous system considered in this paper, while free boundary conditions are preferable in constrained geometries\cite{Yin2,Keesman}.

\section{Calculating the skyrmion lifetime\label{secS5}}

\begin{figure}
\includegraphics[width=\columnwidth]{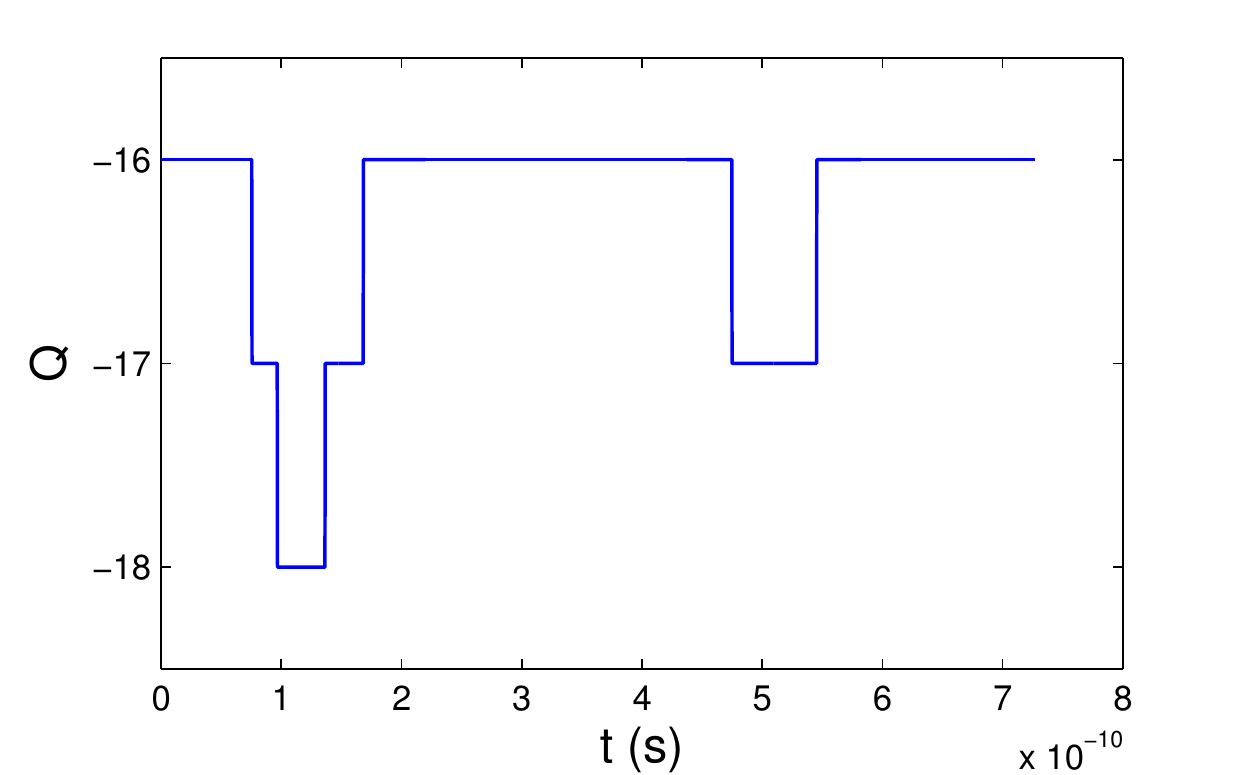}
\caption{(color online) Sample run for determining the skyrmion lifetime showing the topological number $Q$ as a function of time. The parameters are $B=2\,\textrm{T}$, $T=110\,\textrm{K}$, $N=128\times 128$.}
\label{figS4}
\end{figure}

\begin{figure}
\includegraphics[width=\columnwidth]{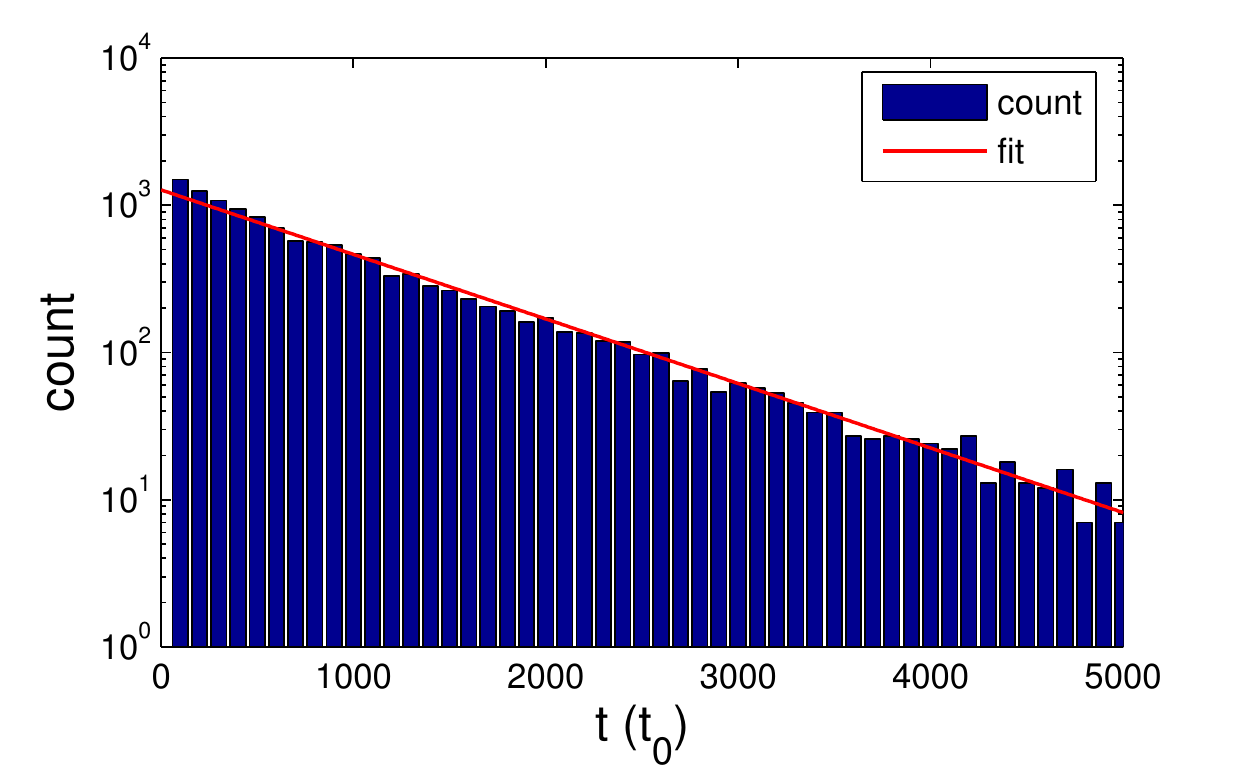}
\caption{(color online) Distribution of time $t$ elapsed between skyrmion creation events. The time unit is $t_{0}=\frac{\hbar}{2\textrm{mRyd}}=2.42\cdot10^{-14}\,\textrm{s}$. As indicated by the fit, the elapsed time follows exponential distribution $\textrm{e}^{-\frac{t}{\tau_{\textrm{cr}}}}$. The parameters are $B=2\,\textrm{T}$, $T=135\,\textrm{K}$, $N=64\times 64$.}
\label{figS5}
\end{figure}

Since we can only measure the topological charge $Q$ of the whole system, the skyrmion lifetime can be calculated from the $Q-t$ diagrams, see Fig.~\ref{figS4}, under some simplifying assumptions. Firstly, we suppose that the increase and decrease of $Q$ is only due to the creation and annihilation of downwards pointing skyrmions, while the presence of upwards pointing skyrmions or antiskyrmions is excluded. This is largely justified at $B=2\,\textrm{T}$ by energy considerations given in Appendix~\ref{secS3}. Even if such structures with opposite topological charge would form in the system, we expect that the energy barrier between the field-polarized state and the skyrmion will mainly depend on the isotropic exchange interactions in the system\cite{Yin,Han}, which makes no difference between skyrmions and antiskyrmions with opposite topological charges.

Secondly, it is impossible to directly follow the creation and annihilation of a single skyrmion if only the net topological charge is calculated. As a seemingly solid approximation we may assume that the time evolution of the skyrmion density per spin $n_{\textrm{sk}}$ follows the master equation
\begin{eqnarray}
\frac{\textrm{d}n_{\textrm{sk}}}{\textrm{d}t}=\frac{1}{\tau_{\textrm{cr}}}-n_{\textrm{sk}}\frac{1}{\tau_{\textrm{sk}}},\label{eqnS8}
\end{eqnarray}
where $\frac{1}{\tau_{\textrm{cr}}}$ is the creation rate of skyrmions due to thermal fluctuations and $\tau_{\textrm{sk}}$ is the skyrmion lifetime. In the stationary case, this simplifies to
\begin{eqnarray}
\tau_{\textrm{sk}}=\tau_{\textrm{cr}}\left<n_{\textrm{sk}}\right>,\label{eqnS9}
\end{eqnarray}
where $\tau_{\textrm{cr}}$ and $\left<n_{\textrm{sk}}\right>$ are directly measurable by calculating the average time between skyrmion creation events and the average skyrmion number, respectively. We note that none of the parameters in Eq.~(\ref{eqnS9}) depend on the lattice size. During the simulations, we can calculate the average number of skyrmions instead of the density, which is expected to scale linearly with the number of spins $N$. However, the average time between skyrmion creation events is inversely proportional to $N$ because the temperature uniformly excites the spins in the simulated area. Indeed, we found that $\tau_{\textrm{sk}}$ calculated as the product of the average time between skyrmion creation events and the average skyrmion number was the same within statistical error for the lattice sizes $N=128\times128$ and $N=64\times64$.

We note that although the topological charge is quantized, the inflation and shrinking of the skyrmion core is not instantaneous but follows a certain time evolution\cite{Verga}, similarly to the formation and absorption of bubbles in boiling liquids. Therefore, it is not likely that the skyrmion lifetime, that is the time difference between the creation and annihilation of the same skyrmion, follows exponential distribution as indicated in Eq.~(\ref{eqnS8}), since it is generally used to describe the distribution of instantaneous events such as the detection of particles in a detector or scattering in a solid. On the other hand, the skyrmion creation events, that is the jumps in $Q\left(t\right)$, are instantaneous, meaning that they likely follow a homogeneous Poisson point process. The distribution between neighboring points in the Poisson process is exponential, and Fig.~\ref{figS5} indicates that the average time between skyrmion creation events is in agreement with this distribution. Fortunately, it is known from the theory of queueing processes that even if the skyrmion lifetime does not follow exponential distribution and therefore Eq.~(\ref{eqnS8}) does not hold, the connection between the expectation values in Eq.~(\ref{eqnS9}) holds as long as the skyrmion creation process is Markovian (a homogeneous Poisson point process), and Eq.~(\ref{eqnS9}) is sufficient for the calculation of the average lifetime. For details about M/G/$\infty$ queueing processes see Ref.~\cite{Newell}.

We emphasize that considering an ensemble of skyrmions based on Eq.~(\ref{eqnS9}) is unavoidable in the fluctuation-disordered state where the average distance between the skyrmions is small. In some recent publications\cite{Yin,Hagemeister}, the skyrmion lifetime was calculated by taking into account a single metastable skyrmion on a ferromagnetic background. Although in this limit the average distance between skyrmions should be considerably larger than in the fluctuation-disordered state, it was shown in Ref.~\cite{Hagemeister} that the calculated skyrmion lifetime depends on the size of the simulated area, and multiple skyrmions may be present simultaneously if the system size is increased. As was discussed after Eq.~(\ref{eqnS9}), we did not find such a dependence of the skyrmion lifetime on the system size. Furthermore, the presence of an ensemble of skyrmions opens different types of processes for a change in the topological charge beyond the usual nucleation and annihilation on a homogeneous background. These processes include the merging of two skyrmions into one, and the splitting of a single skyrmion into two. Similar merging effects have been discussed in Ref.~\cite{Milde} for a three-dimensional system, when the skyrmion lattice transforms into the spin spiral state as the external magnetic field is decreased.

Finally, we note that we only performed the simulations for a fixed value of the damping parameter, $\alpha=0.05$. It is known that the skyrmion lifetime should depend on the damping\cite{Verga}, but the energy barrier $\Delta E$ in Eq.~(\ref{eqn9}) between the skyrmion and field-polarized state should not, as it is related to the interactions in the spin model. Due to the exponential dependence on $\Delta E$, its value is significantly more important during the determination of lifetime and critical temperature than the value of the prefactor $\tau_{0}$.

\end{document}